%% file: dipole.tex
\pdfoutput=1
\documentclass[11pt, reqno,preprint]{article}
\usepackage{jheppub}
\usepackage{epsfig}
\usepackage{amssymb}
\usepackage{amsmath}
\usepackage{mathrsfs}
\usepackage{hyperref}
\usepackage{multirow}

\def\be{\begin{equation}}
\def\ee{\end{equation}}
\def\ba{\begin{eqnarray}}
\def\ea{\end{eqnarray}}
\def\nl{\nonumber\\}

\def\b{\beta}

\def\Li{\textrm{Li}}
\def\l{\langle}
\def\r{\rangle}
\def\SS{\mathcal{S}}

\def\eps{\epsilon}

\def\UU{\mathcal{U}}

\def\z#1#2{\alpha_{{#1}{#2}}}
\def\zsq#1#2{\alpha_{{#1}{#2}}^2}
\def\s#1#2{s_{{#1}{#2}}}
\def\ss#1{s_{#1}}
\def\ssl#1#2{s_{(#1){#2}}}
\def\ssr#1#2{s_{{#1}(#2)}}

\def\br#1#2#3{[{#1}\,{#2}\,{#3}]}
\def\Fa#1#2#3{F_{[{#1}\,{#2}\,{#3}]}}
\def\Fb#1#2#3{F_{[{#1}\,{#2}\,{#3}]}}
\def\Fc#1#2#3{F_{[{#1}\,{#2}\,{#3}]}}
\def\Fsuba#1#2#3{F^{\rm sub}_{[{#1}\,{#2}\,{#3}]}}
\def\Fsubb#1#2#3{F^{\rm sub}_{[{#1}\,{#2}\,{#3}]}}
\def\Fsubc#1#2#3{F^{\rm sub}_{[{#1}\,{#2}\,{#3}]}}
\def\Q#1#2#3{Q_{[{#1}\,{#2}\,{#3}]}}
\def\Qsq#1#2#3{Q^2_{[{#1}\,{#2}\,{#3}]}}

\def\tv{\underline{v}}
\def\brv#1#2#3{\{#1\,#2\,#3\}}
\def\gm#1#2#3{\Gamma_{\{#1\,#2\,#3\}}}

\newcommand{\lsim}{\mathrel{\hbox{\rlap{\lower.55ex \hbox{$\sim$}} \kern-.3em \raise.4ex \hbox{$<$}}}}
\newcommand{\gsim}{\mathrel{\hbox{\rlap{\lower.55ex \hbox{$\sim$}} \kern-.3em \raise.4ex \hbox{$>$}}}}

\def\eps{\epsilon}

\def\ren{{\rm ren}}
\def\xb{\bar{x}}

\title{High-Energy Evolution To Three Loops}

\author[a]{Simon Caron-Huot}
\author[a, b, c]{and Matti Herranen}
\affiliation[a]{Niels Bohr International Academy and Discovery Center, Blegdamsvej 17, Copenhagen 2100, Denmark}
\affiliation[b]{Department of Physics, University of Jyv\"askyl\"a, P.O. Box 35 (YFL), FI-40014
University of Jyv\"askyl\"a, Finland}
\affiliation[c]{Helsinki Institute of Physics, P.O. Box 64, FI-00014, Helsinki, Finland}
\emailAdd{schuot@nbi.dk, matti.h.herranen@jyu.fi} \abstract{
The Balitsky-Kovchegov equation describes the high-energy growth of gauge theory scattering amplitudes
as well as nonlinear saturation effects which stop it.
We obtain the three-loop corrections to the equation in planar $\mathcal{N}=4$ super Yang-Mills theory.
Our method exploits a recently established equivalence with the physics of soft wide-angle radiation, so-called non-global logarithms,
and thus yields at the same time the three-loop evolution equation for non-global logarithms.
As a by-product of our analysis, we develop a Lorentz-covariant method to subtract infrared and collinear divergences
in cross-section calculations in the planar limit.  We compare our result in the linear regime
with a recent prediction for the so-called Pomeron trajectory,
and compare its collinear limit with predictions from the spectrum of twist-two operators.
}

\notoc

\begin{document}

\maketitle

\section{Introduction}

In high-energy scattering, the aspect of a particle depends on the energy scale at which it is probed.
In hadronic collisions
this effect can be seen in the well known energy dependence of parton distribution functions. 
The energy dependence can be accessed in a more detailed way by looking at less inclusive observables,
for example ones probing correlations between very different rapidities, opening a window on the transverse structure of the projectile.
One then encounters another fundamental evolution equation of QCD,
the Balitsky-Fadin-Kuraev-Lipatov (BFKL) equation \cite{Kuraev:1977fs,Balitsky:1978ic}.

In contrast to other evolution equations, which are typically linear,
nonlinear effects can also play a role in rapidity evolution:
once scattering at a given impact parameter has reached opacity, it must saturate.
A nonlinear evolution equation which
incorporates such effects within perturbation theory has been derived by Balitsky and Kovchegov \cite{Balitsky:1995ub,Kovchegov:1999yj}.
Asymptotically, saturation may occur at distances shorter than the nonperturbative
scale $\Lambda_{\rm QCD}^{-1}$, justifying the use of perturbation theory \cite{McLerran:1993ni,Gelis:2010nm}.
For many observables, such as inclusive jet correlations or deep inelastic scattering,
perturbation theory is also justified by the large momentum transfer in the problem (see for example \cite{Caporale:2015vya,Iancu:2015joa} and references therein).
The need to control higher order corrections, and the need to better understand the theory at finite coupling,
motivate a deeper look into the perturbative series.

The next-to-leading-order evolution equation has been known for some time \cite{Balitsky:2008zza}.
It reproduces, in the appropriate limit, the next-to-leading order BFKL Pomeron trajectory \cite{Fadin:1998py,Ciafaloni:1998gs}.
A notable feature is that the degree of nonlinearity and its complexity increases with each new order in perturbation theory.
This is a rather unfamiliar situation which makes it unclear how to best formulate the equation at finite coupling.
Furthermore, the corrections have turned out to be numerically large. This has
been attributed to collinear effects, suggesting a possibility to resum them at higher orders at both the linear (BFKL)
and nonlinear level \cite{Salam:1998tj,Vera:2005jt,Iancu:2015vea}. 
In order to shed light on these issue, and to critically assess the quality of proposed resummations, higher-loop data is clearly highly desirable.

The aim of this paper is to initiate a systematic study of the Balitsky-Kovchegov and BFKL equations at three loops and beyond.
Specifically, as a first step, we will derive its three-loop (next-to-next-to-leading order) correction in the planar limit of $\mathcal{N}=4$ super Yang-Mills (SYM).
This calculation is made possible by recent conceptual and technological developments in the calculation of scattering amplitudes.
Our methods remain however essentially diagrammatic and we expect them to prove applicable to QCD in a next step.

The SYM model is an ideal stepping stone for several reasons. First, partial cross-checks are available
due to a recent and highly remarkable prediction of the Pomeron trajectory exploiting integrability in this model \cite{Gromov:2015vua,Velizhanin:2015xsa}.
Such tests are valuable both from the perturbative and integrability perspective.
At the nonlinear level, the interactions to be predicted
are related to structure constants \cite{Balitsky:2015tca}, soon to be within reach of similar methods.
Together with the AdS/CFT correspondence at strong coupling \cite{Brower:2006ea},
these hint at a possible exact description of the Pomeron and its interactions at finite coupling in this model.

\subsection{High-energy scattering, soft gluons, and non-global logarithms}

A modern description of high-energy forward scattering is based on the eikonal approximation:
fast projectiles and targets are approximated by null Wilson lines $U$. More precisely, by a collection
of such Wilson lines, reflecting the transverse structure of the colliding objects at the given rapidity scale \cite{Balitsky:1995ub}.
It is simple to translate this language to that of classic Regge theory:
the reggeized gluon is the state sourced by (the logarithm of) a null Wilson line \cite{Caron-Huot:2013fea}.

The three-loop calculation in this paper is enabled by a recently established correspondence with the physics
of wide-angle soft radiation, sometimes called ``non-global logarithms.''
Consider the QCD decay of a color singlet state like a virtual photon or $Z$ boson, with energy $Q$.
A representative observable, sensitive to soft wide-angle radiation, is the probability to \emph{not} find radiation with energy above a cutoff $\mu$ within an exclusion region $R$ (see fig.~\ref{fig:twodipoles}).
If the cutoff $\mu$ is low, this probability is small
and controlled in the planar approximation ('t Hooft limit $N_c\to\infty$) by the Banfi-Marchesini-Smye evolution equation \cite{Banfi:2002hw}:
\be
\frac{d}{d\log\mu}U_{12} =
\frac{\lambda}{16\pi^2}
\int \frac{d\Omega_0}{4\pi} \frac{\z{1}{2}}{\z{1}{0}\z{0}{2}}\big(2U_{12}-2U_{10}U_{02}\big)\equiv \frac{\lambda}{16\pi^2} K^{(1)}U_{12}. \label{K1intro}
\ee
This resums large logarithms $\log\frac{Q}{\mu}$.
Here $\z{i}{j}\equiv\frac{1{-}\cos\theta_{ij}}{2}$, and the subscripts denote the angles of outgoing partons;
the dipole
$U_{ij}=\frac{1}{N_c}{\rm Tr}[U(\theta_i) U^\dagger(\theta_j)]$ is a function of two angles which
can be interpreted (see below) as the trace of a color dipole at angles $\theta_i$ and $\theta_j$.

The basic physics of this equation is that the color flow, and therefore the energy flow,
is affected by radiation of an extra gluon at angle $\theta_0$.
The observable, through the exclusion region $R$, is encoded by the infrared boundary condition
that $U_{ij}=0$ when either $i$ or $j$ are in $R$. Qualitatively, the evolution leads to an increased effective size of the exclusion region,
as radiation near the allowed boundaries become more and more in danger of leaking out.%
\footnote{
The form (\ref{K1intro}) is valid provided that $R$ is smooth enough that no jets are forced to be narrow.  This is assumed here in order to avoid further subtractions of collinear singularities as in the original setup \cite{Banfi:2002hw}, thereby focusing on soft wide-angle radiation
and preserving the most symmetrical form of the equation.
}

This equation is mathematically equivalent to the Balitsky-Kovchegov equation, which governs the rapidity dependence of perturbative high-energy scattering near the forward direction.  In this context, the trivial fixed point $U_{ij}{=}1$ represents a transparent target, which is unstable:
by linearizing in the departure $(1{-}U)$, which gives the BFKL equation, one finds
a growing solution known as the BFKL Pomeron.  The nonlinear term then
accounts for a class of saturation effects which stop the growth (locally in the transverse plane) toward the attractive, opaque,
fixed-point $U_{ij}{=}0$.

The nonlinear term in both equations share a similar physical origin: in both cases
one is interested in the probability that something does \emph{not} happen, while many possibly complicated things \emph{may} happen \cite{Marchesini:2015ica}. Indeed,
to describe the probability to \emph{not} radiate in a certain region, one must keep track of all \emph{allowed} radiation,
which is what the nonlinear term of eq.~(\ref{K1intro}) produces.  Similarly, in near-forward scattering, one measures the probability for a projectile to \emph{not} be destroyed at a given impact parameter.
The two evolutions share other physical similarities: both are dominated by  soft gluons, and both feature ``opaque'' and ``transparent'' regimes.

Given these similarities, it seems natural to expect a relationship between these two problems.
The geometry is however different.
To establish a rigorous map turns out to require a conformal transformation \cite{Hatta:2008st,Caron-Huot:2015bja},
which equates detector measurements at infinity with the physics of a fast particle crossing a Lorentz-contracted target (also known as a shockwave).
This had been used notably by Hofman and Maldacena and others
to describe detector measurements in conformal field theories \cite{Hofman:2008ar,Hatta:2008st,Cornalba:2009ax} and at the same time gain new insight into high-energy scattering.
This conformal transformation is just the stereographic projection of a two-sphere onto the
transverse impact parameter plane:
\be
 \int\frac{d\Omega}{4\pi} \Leftrightarrow \int \frac{d^2z}{\pi}, \qquad \frac{1-\cos\theta_{ij}}{2} \Leftrightarrow m^2|z_i-z_j|^2,\qquad
 \frac{d}{d\log\mu} \Leftrightarrow -\frac{d}{d\eta}.
 \label{stereographic_projection}
\ee
Here $m$ is an arbitrary mass scale and $\eta$ is rapidity.
Under this dictionary, the Banfi-Marchesini-Smye equation (\ref{K1intro})
becomes precisely the Balitsky-Kovchegov equation, as was noted early on \cite{Weigert:2003mm}.

In this paper we will exploit this correspondence and work exclusively on the non-global logarithm side,
which is technically advantageous due to a body of knowledge on the infrared and collinear factorization of amplitudes and cross-sections.
This correspondence was emphasized and tested explicitly at two-loops in \cite{Caron-Huot:2015bja}, where the full two-loop BFKL/BK equation (including running coupling effects and non-planar corrections) was re-derived starting from non-global logarithm problem.

The evolution equation (\ref{K1intro}) at finite coupling is best viewed
as a renormalization group (RG) equation:
\be
 \left[\frac{d}{d\log\mu} +\beta(g^2) \frac{d}{dg^2} -K\right] \sigma[U;\mu]=0, \label{RG}
\ee
where the  \emph{color density matrix}, or \emph{weighted cross-section}  $\sigma[U]$,
is defined operationally by weighing each final state parton by a color rotation $U(\theta_i)$ \cite{Caron-Huot:2015bja} (see also \cite{Nagy:2012bt}).
These color rotations can be understood as Wilson lines $U(\theta_i$) accounting for the effect of more infrared radiation
(these Wilson lines connect the decaying state in the matrix element and its conjugate) \cite{Larkoski:2015zka,Becher:2015hka}.
In the planar limit, the color factors reduce to products of color dipoles and the color density matrix simplifies to a single
function $U_{ij}$ of only two angles,
as shown in fig.~\ref{fig:twodipoles}, which illustrates the ``$UU$" term in eq.~(\ref{K1intro}).

In both eqs.~(\ref{K1intro}) and (\ref{RG}), $\mu$ is an infrared cutoff below which all radiation is inclusive.
In our practical calculation we will work within dimensional regularization to $D=4{-}2\eps$ ($\eps<0$).
Then the cutoff $\mu$ appears in a renormalization procedure. Following standard procedure, this is equivalent to
integrating the RG equation from the deep infrared:
\be
 \sigma^{\rm bare}[U]= \mathcal{P}\exp\left[-\int_0^\mu\frac{d\lambda}{\lambda} K(g^2(\lambda))\right]\sigma^{\rm ren}[U;\mu]\,,
 \label{integratedRG}
\ee
where, writing $g^2(\lambda)=g^2(\mu)(\lambda/\mu)^{-2\eps}+O(g^4)$ for the running coupling in $D$ dimensions,
one can see that the integral produces $1/\eps$ poles.
The subtraction then cancels the poles in the bare amplitude so as to make $\sigma^{\ren}[U;\mu]$ finite as $\eps{\to}0$. That the divergences exponentiate in precisely this way
was proved to all orders in ref.~\cite{Caron-Huot:2015bja}, exploiting known results on the factorization of soft partons \cite{Catani:1999ss,Feige:2014wja}.
The upshot of eq.~(\ref{integratedRG}) is that we can use the $1/\eps$ poles in the dimensionally regulated weighted cross-section to read off the
non-global-logarithm/Balitsky-Kovchegov kernel $K$.  Note that this is identical to the standard procedure to extract (ultraviolet) anomalous dimensions of local operators, by using their $1/\eps$ poles.
The fact that divergences (either infrared or ultraviolet) are controlled by renormalization group equations
is of course due to the Wilsonian decoupling between physics at different scales.

This paper is organized as follows. In section \ref{sec:notations} we introduce useful notations for soft currents and phase space integrals.
In section \ref{sec:two-loops} we revisit the two-loop calculation, improving on previous treatments by introducing a scheme where Lorentz symmetry
is manifest at each step; under the correspondence with the Regge limit, this is equivalent to maintaining the conformal symmetry of the BK equation.
In section \ref{sec:three-loops} we perform the three-loop calculation, paying special attention to the combinatorics of subdivergences
and their cancellations, culminating in the final result for the nonlinear equation in subsection \ref{ssec:final}.  In section \ref{sec:linear} we analyze its linearized limit,
compute its eigenvalues, and compare it with integrability predictions. Finally section \ref{sec:conclusion} contains our concluding remarks.
In three appendices, we record the one-loop double soft current squared (appendix \ref{app:sixpoint}), we detail
our algorithm to compute finite angular or transverse integrals (appendix \ref{app:integrals}), and record the three-loop eigenvalue (appendix \ref{app:eigen}).

\section{Notations} \label{sec:notations}

\begin{figure}
        \centering
\centering \def\svgwidth{12cm}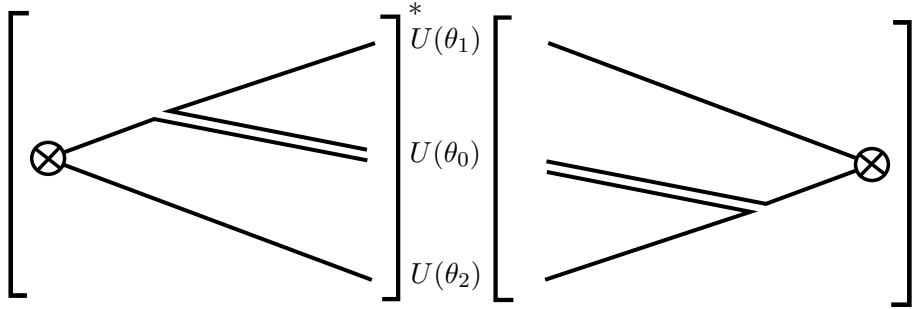
\caption{Soft wide-angle radiation: radiation is allowed in some region but excluded in another.
To keep track of the allowed radiation we use a color density matrix, defined by applying an angle-dependent
color rotation $U(\theta_i)$ between the matrix element and its conjugate for each final state particle.
In the planar limit this configuration reduces to a product of two color dipoles.
}
\label{fig:twodipoles}
\end{figure}

The calculation of $K$ requires squared matrix elements for emitting soft partons off two color-correlated parents (``dipole''),
the so-called soft currents.
The evolution equation, just like the soft currents, is universal and does not depend on details of the underlying short-distance process, only on the color charges and angles
of the outgoing partons.
Final states are then weighted, in the planar limit, by a product of color dipoles (see fig.~\ref{fig:twodipoles}).
For each such product, it is useful to pull out a universal factor
which accounts for its dimensionality and most singular limits.
We thus write the contribution from the soft current with $n$ soft partons to the $n$-loop cross-section, starting from a parent dipole $U_{12}$ along directions $p_1$ and $p_2$, as:
\begin{subequations}
\ba
 \sigma^{(1)}_1 &=& \int_{p_0} \frac{\s{1}{2}}{\s{1}{0}\s{0}{2}}\,2\, U_{10}U_{02} \,\Fa{1}{0}{2}\,, \label{real_sigma_1}
\\
 \sigma^{(2)}_2 &=& \int_{p_0,p_{0'}} \frac{\s{1}{2}}{\s{1}{0}\s{0}{0'}\s{0'}{2}}\,4\,U_{10}U_{00'}U_{0'2}\,\Fb{1}{00'}{2}\,, \label{real_sigma_2}
 \\
 \sigma^{(3)}_3 &=& \int_{p_0,p_{0'},p_{0''}} 
\frac{\s{1}{2}}{\s{1}{0}\s{0}{0'}\s{0'}{0''}\s{0''}{2}}\,8\,U_{10}U_{00'}U_{0'0''}U_{0''2}\,\Fc{1}{00'0''}{2}\,,\qquad \mbox{etc.} \label{real_sigma_3}
\ea
\label{real_sigma}
\end{subequations}
\!\!Here loops are counted in powers of $g^2\equiv \frac{g^2_{\rm YM}N_c}{16\pi^2}=\frac{\alpha_s N_c}{4\pi}$,
$U_{ij}=\frac{1}{N_c}{\rm Tr}[U(\theta_i) U^\dagger(\theta_j)]$,
and phase-space integrals are normalized as
$\int_{p_0} \equiv 16\pi^2\int \frac{\mu^{2\eps}d^{3{-}2\eps}p_0}{(2\pi)^{3{-}2\eps}2p_0^0}$.
For the Mandelstam invariants and their multi-index generalizations, we let:
\be
\s{i}{j}=2p_i{\cdot}p_j, \qquad \ssr{i}{jk}=2p_i{\cdot}(p_j+p_k),
\qquad \ss{ijk}=2(p_i{\cdot}p_j+p_i{\cdot}p_k+p_j{\cdot}p_k)\,.
\ee
All invariants will always be positive (timelike), since we assume a color singlet initial state.
Naturally for our setup, on-shell momenta will be split into an  energy $a_0$ and angular parts $\beta_0$:
$p_0^\mu=a_0\beta_0^\mu$ where $\beta_0^\mu=(1,\vec{n}_0)$ is null. The Lorentz-invariant phase space
measure correspondingly splits into an energy and angular parts:
\be
\int_{p_0}= \mu^{2\eps}\int_0^\infty 2da_0(2a_0)^{1{-}2\eps}\times\int_{\beta_0}\,,
\qquad \int_{\beta_0} \equiv \int\frac{d^{2{-}2\eps}\Omega_0}{(4\pi)^{1{-}2\eps}}. \label{energy_and_angles}
\ee
For angles we write $\z{i}{j}=\frac{1-\cos\theta_{ij}}{2}$, which runs between 0 and 1.
Throughout, we will use the subscripts $0$, $0'$, $0''$ to index radiated gluons.

The various factors of $2$ in our definitions
have been chosen to simplify limits and preclude unnecessary $(\log2)$'s in integrated expressions.
For example, for one soft gluon, $F[1\,0\,2]$ is the square of the well-known eikonal soft current.
Including the factor $T^aT^a/N_c=1/2$ from the color sum, this  evaluates to
\be
 \frac{2\s{1}{2}}{\s{1}{0}\s{0}{2}}\Fa{1}{0}{2} \equiv \frac12\left| \frac{p_1^\mu}{p_1\cdot p_0}-\frac{p_2^\mu}{p_2\cdot p_0}\right|^2
\longrightarrow \Fa{1}{0}{2} = 1\,. \label{F1}
\ee
For two soft partons one needs the square of the double soft current, described for example in \cite{Catani:1999ss}.
The result after squaring it and including the fermions and scalars of $\mathcal{N}=4$ SYM
can be borrowed from formulas of ref.~\cite{Caron-Huot:2015bja} (section 3), also rederived below in subsection \ref{ssec:triplesoft}:
\be
 \Fb{1}{00'}{2}=
   1+ \frac{\s{1}{2}\s{0}{0'}+\s{1}{0}\s{0'}{2}-\s{1}{0'}\s{0}{2}}{2\ssr{1}{00'}\ssl{00'}{2}}\,. \label{F2}
\ee
One can easily verify that this factorizes in soft limits:
\be
\Fb{1}{00'}{2}\xrightarrow{|p_0| \ll |p_{0'}|} \Fa{1}{0}{0'}\Fa{1}{0'}{2} = 1,\qquad
\Fb{1}{00'}{2}\xrightarrow{|p_{0'}| \ll |p_{0}|} \Fa{0}{0'}{2}\Fa{1}{0}{2} = 1\,. \label{softF2}
\ee
Our three-loop computation builds on the one-loop corrections to $\Fb{1}{00'}{2}$
and the tree-level three-parton amplitude $\Fc{1}{00'0''}{2}$, which will be efficiently obtained
as described below.

\section{Two-loop evolution: fixing a convenient scheme}
\label{sec:two-loops}

The next-to-leading order correction to the Balitsky-Kovchegov equation was obtained in QCD and $\mathcal{N}=4$ SYM in \cite{Balitsky:2008zza,Balitsky:2009xg}.
It was postulated \cite{Avsar:2009yb} and verified explicitly
that the same kernel governs non-global logarithms \cite{Caron-Huot:2015bja}.

In the latter reference, soft partons were organized in terms of their energies. Because ``energy'' is not Lorentz invariant,
this scheme did not manifest Lorentz invariance, which had to be restored manually through a finite renormalization
(guaranteed to exist given the Lorentz invariance of the underlying theory).
This is formally similar to the transformation used to reach to the so-called conformal scheme in the BK literature \cite{Balitsky:2009xg}.
Indeed the mapping (\ref{stereographic_projection}) interchanges the Lorentz and conformal symmetry of the two problems.

Here we improve on this by using explicitly Lorentz-invariant cutoffs.
To fully define the scheme in which our three-loop result will apply,
we thus quickly revisit the two-loop calculation.

\subsection{One-loop in Lorentz-invariant form}

The idea is to define the evolution so that its exponentiation
generates the emission probability of a soft gluon, in the soft approximation,
integrated over a complete phase space region bounded by a Lorentz-invariant cutoff.
For example, at one-loop, we define the anomalous dimension $K^{(1)}$ so that its integral
(following the first term in the expansion of eq.~(\ref{integratedRG})) matches the 
emission amplitude given in eqs.~(\ref{real_sigma_1}),(\ref{F1}):
\be
 -\int_0^\mu \frac{d\lambda}{\lambda}\lambda^{-2\eps} K^{(1)}U_{12} \equiv \int_{p_0} \theta(\Q{1}{0}{2}<\mu)\frac{s_{12}}{s_{10}s_{02}}
 \big(2U_{10}U_{02}-2U_{12}\big)\,,
\label{K1def}
\ee
where $\theta(x<y)$ is a step function forcing $x$ to be smaller than $y$,
and $\Qsq{1}{0}{2}\equiv \frac{s_{10}s_{02}}{s_{12}}$ defines our cutoff.
From this definition one can see that $\Q{1}{0}{2}$ is proportional to the energy of the radiated gluon.
Physically, $\Q{1}{0}{2}$ is the absolute value of its transverse momentum in a frame where the parents $p_1$ and $p_2$ are back to back.
(This ordering variable has been used in many other contexts, see for example \cite{Angeles-Martinez:2016dph}.)
It is the only Lorentz invariant scale that depends on the direction but not the energies of the parent partons.

To find $K^{(1)}$ from the definition (\ref{K1def}), we simply identify the integration over the
energy component of $p_0$ (called $a_0$ in eq.~(\ref{energy_and_angles})) with that over the ordering scale $\lambda$.
More precisely, $\lambda$ is proportional, but not equal, to the energy $a_0$, because of the angle dependence of $\Q{1}{0}{2}$:
\be
\lambda=\Q{1}{0}{2} =2a_0\sqrt{\frac{\z{1}{0}\z{0}{2}}{\z{1}{2}}}\,.
\ee
Inserting this change of variable into the right-hand-side of eq.~(\ref{K1def}) using the measure (\ref{energy_and_angles}),
and stripping off $\int \frac{d\lambda}{\lambda} \lambda^{-2\eps}$ on both sides, we thus get:
\be
 K^{(1)}U_{12} = \int_{\beta_0} \left(\frac{\z{1}{2}}{\z{1}{0}\z{0}{2}}\right)^{1{-}\eps} \big(2U_{12}-2U_{10}U_{02}\big)\,. \label{K1}
\ee
This of course reproduces the one-loop Banfi-Marchesini-Smye equation recorded in (\ref{K1intro}), \emph{except} for the $\eps$ in the exponent,
which arose because of the angular dependence of the ordering variable $\Q{1}{0}{2}$.
This exponent ensures exact Lorentz invariance in any dimension, not only in the $\eps\to0$ limit\footnote{An angular integral $\int d^{2{-}2\eps}\Omega_0 \,I(\b_0)$ is Lorentz invariant if $I$ is homogenous of degree $-(2{-}2\eps)$ in $\b_0$. This condition ensures that the rescaling of $\b_0$ under a boost cancels against the Jacobian of the transformation.}, which is critical to ensure Lorentz invariance of the higher-loop corrections to $K$ \cite{Caron-Huot:2015bja}.

We briefly comment on the inclusion of virtual corrections, which simply add the $(-2U_{12})$ term to eq.~(\ref{K1def}). This form is determined by the Kinoshita-Lee-Nauenberg (KLN) theorem \cite{Kinoshita:1962ur,Lee:1964is}, which states that there can be no infrared divergences in a fully inclusive cross-section. This implies, in particular, that $U_{ij}=1$ is a fixed point of the evolution. At any loop order this can (and will) be used to obtain the coefficient of $U_{12}$ from that of other color structures.

\subsection{Lorentz invariant slicing of multi-particle phase spaces}

To move on to higher loops, we define, similarly, scales for multiple emissions:
\be
 \Qsq{1}{0}{2}= \frac{\s{1}{0}\s{0}{2}}{\s{1}{2}},\quad
 \Qsq{1}{00'}{2} \equiv \left(\frac{\s{1}{0}\s{0}{0'}\s{0'}{2}}{\s{1}{2}}\right)^{1/2},\quad
 \Qsq{1}{00'0''}{2} \equiv \left(\frac{\s{1}{0}\s{0}{0'}\s{0'}{0''}\s{0''}{2}}{\s{1}{2}}\right)^{1/3},\quad \mbox{etc.}
\ee
Similar combinations appeared already in the integration measures in eqs.~(\ref{real_sigma}).
The exponents may appear unwieldy, but in practice these definitions will be very convenient
because the scales of complicated processes are equal to appropriate geometric means of subprocess scales,
for example:
\be \Qsq{1}{00'}{2}=\Q{0}{0'}{2}\Q{1}{0}{2}=\Q{1}{0}{0'}\Q{1}{0'}{2}.
\ee
As an organizing principle, when writing the higher-loop contributions to the evolution kernel $K$,
we make sure to completely cover the multi-parton phase space up to a cutoff in $Q$.
Let us consider for illustration a term arising from an $\ell$-loop virtual correction to the emission of two real partons 
($\ell=0$ being the relevant case for the two-loop kernel to be detailed shortly).
If $I^\ell$ denotes the corresponding soft current, the following expression integrates it over all the phase space with $\Q{1}{00'}{2}$ below the cutoff:
\be
 \int_{p_0,p_{0'}} \frac{\s{1}{2}}{\s{1}{0}\s{0}{0'}\s{0'}{2}}\,\theta(\Q{1}{00'}{2}<\mu)\,I^\ell(p_0,p_{0'}).
\label{two_body}
\ee
Importantly, the integrand will always be homogeneous, due physically to the fact that the Yang-Mills
coupling is dimensionless.
More precisely, within dimensional regularization,
the $\ell$-loop correction to the two-parton emission has an overall dimension determined by the running coupling
$g^2(\lambda)\approx \lambda^{-2\eps}$, raised to the power $(\ell{+}2)$.
We thus change variable from the two energies $(a_0,a_{0'})$ to the overall scale
$\lambda\equiv\Q{1}{00'}{2}$ and relative energy $\tau=a_0/a_{0'}$.
Dimensional reasoning then implies that, after factoring out the running coupling
evaluated at that scale $\Q{1}{00'}{2}$, the integrand becomes homogeneous and depends only on the ratio $\tau$, but not $\lambda$:
\be
 I^\ell(p_0,p_{0'}) = (g^2(\Q{1}{00'}{2}))^{\ell+2} \times \tilde{I}^\ell(\tau \beta_0,\beta_{0'})\,.
\ee
With this change of variable the two-particle phase-space then factors as
\be
\mbox{eq.~(\ref{two_body})} =
 \int_0^\mu \frac{d\lambda}{\lambda}
 (g^2(\lambda))^{\ell{+}2}
 \int_{\b_0,\b_{0'}}
 \left(\frac{\z{1}{2}}{\z{1}{0}\z{0}{0'}\z{0'}{2}}\right)^{1{-}\eps}
 \int_0^\infty \frac{d\tau}{\tau} \,\tilde{I}(\tau \b_0,\b_{0'}),
\label{phase_space_2_partons}
\ee
The integral over the scale $\lambda$ precisely matches what appears in the integrated renormalization group equation (see eq.~(\ref{integratedRG})),
so by simply stripping it off we get the contribution to the kernel $K^{(\ell{+}2)}$, simply generalizing eq.~(\ref{K1}).
Note, importantly, that the equality is exact to all orders in $\eps$ and holds not only for the $1/\eps$ poles.  The only assumption
is that the integrand $I^\ell$ is computed in the leading soft approximation where
the energy scales of the parent partons does not enter, e.g. $I^\ell$ is the standard soft current comes from soft currents, 
The generalization to more partons is immediate: in our slicing scheme we will always get integrations that depend only
over relative energies $\tau$, with an $\eps$-free measure.
(Strictly speaking, the identity (\ref{phase_space_2_partons}) is only valid when the integrand $I^\ell$ vanishes in its factorization limits $\tau\to 0,\infty$ so the $\tau$-integral converges,
which holds for the subtracted integrand $F^{\rm sub}$ to be defined shortly.)


This equivalence between scale integrals and energy integrals can also be applied in the reverse direction,
to subtract the iteration of the lower-loop kernels generated by the path-ordered exponential (\ref{integratedRG}).
For example the product of two $K^{(1)}$'s corresponding to the successive emission of parton $0$ between 1 and 2,
followed by parton $0'$ between $0$ and $2$, can be written as
\be\hspace{-1mm}
\mu^{4\eps} \int_0^\mu \frac{d\lambda}{\lambda}\lambda^{-2\eps}
  \int_0^\lambda \frac{d\lambda'}{\lambda'}\lambda'^{-2\eps} \int_{\beta_0,\beta_{0'}}\!\!\!
r_{[0\,0'\,2]}r_{[1\,0\,2]}
  = \int_{p_0,p_0'}  \frac{\s{0}{2}}{\s{0}{0'}\s{0'}{2}}\frac{\s{1}{2}}{\s{1}{0}\s{0}{2}}\theta(\Q{0}{0'}{2}<\Q{1}{0}{2}<\mu)\,, \label{iterationK1}
\ee
where $r_{[1\,0\,2]}=\left(\z{1}{2}/(\z{1}{0}\z{0}{2})\right)^{1{-}\eps}$ is the angular measure in eq.~(\ref{K1}).

\subsection{Quick rederivation of two-loop evolution}
\label{ssec:2loop}

With this technology it is now rather straightforward to re-derive the two-loop evolution equation.
Let us start with the contribution from two real partons.
Matching with eq.~(\ref{integratedRG}), this requires the squared matrix element for two partons, minus the iteration of one-loop subprocesses.
This later subtraction will neatly remove all subdivergences. There are two possible one-loop subprocesses: either $p_0$ or $p_0'$ can be radiated first.
The relation (\ref{iterationK1}) allows to subtract these directly at the integrand level, by defining a subtracted soft current:
\be
F^{\rm sub}_{[1\,00'\,2]} \equiv \Fb{1}{00'}{2} 
-\theta\big(\Q{0}{0'}{2}{<}\Q{1}{0}{2}\big)
-\theta\big(\Q{1}{0}{0'}{<}\Q{1}{0'}{2}\big). \label{F2sub}
\ee
Multiplying with the product of dipoles $U_{10}U_{00'}U_{0'2}$ in eq.~(\ref{real_sigma_2}) and
removing the overall scale integral using eq.~(\ref{phase_space_2_partons}), we get $K$ as a convergent integral:
\be
 K^{(2)}U_{12}= \int_{\b_0,\b_{0'}} \left(\frac{\z{1}{2}}{\z{1}{0}\z{0}{0'}\z{0'}{2}}\right)^{1{-}\eps}
 \int_0^\infty\frac{d\tau}{\tau}
 (-2U_{10}U_{00'}U_{0'2}) 2F^{\rm sub}_{[1\,(\tau\b_0)\b_{0'}\,2]} + \ldots
 \label{K2real0}
\ee
where the omitted terms involve virtual corrections (involving products of fewer than three $U$ dipoles).
The $\tau$ integral converges absolutely in both the $\tau\to0$ and $\tau\to\infty$ limits
thanks to the factorization of the soft current $F$ noted in eq.~(\ref{softF2}).

Note that we have omitted the $\mu$ upper cutoff in the step functions in $F^{\rm sub}_{[1\,00'\,2]}$.
This is because all the $1/\eps$ poles at two loops come from the infrared region where $p_0\sim p_{0'}\ll \mu$,
where this cutoff plays no role \cite{Caron-Huot:2015bja}. The region near the upper cutoff only affects the two-loop amplitude by a finite amount,
thus affecting the evolution starting only from three loops (in a way which can be systematically accounted for, see eq.~(\ref{Fsub3})).
The $\tau$-integral in (\ref{K2real0}), using the explicit expression (\ref{F2}), involves only elementary integrals
and gives a simple angular function
\be
 K^{(2)}_{[1\,00'\,2]} \equiv  \int_0^\infty\frac{d\tau}{\tau} 2F^{\rm sub}_{[1\,(\tau\b_0)\b_{0'}\,2]} = 2\log \frac{\z{1}{2}\z{0}{0'}}{\z{1}{0'}\z{0}{2}} + \left(1+\frac{\z{1}{2}\z{0}{0'}}{\z{1}{0}\z{0'}{2}{-}\z{1}{0'}\z{0}{2}}\right)\log\frac{\z{1}{0}\z{0'}{2}}{\z{1}{0'}\z{0}{2}}\,. \label{K2real}
\ee
Finally we turn to the virtual corrections, which can have color factors $U_{10}U_{02}$ or $U_{12}$. They are strongly constrained by physical principles:
Lorentz invariance, the absence of collinear singularities, and the KLN theorem.  A simple way to solve these
constraints is to add $(U_{10}U_{02}+U_{10'}U_{0'2})$ to the color factor in eq.~(\ref{K2real0}),
which automatically removes collinear singularities when $0{\parallel}0'$ (where $U_{00'}\to 1$)
and fulfills KLN. By Lorentz invariance, the remainder is then determined up to a single multiple of one-loop:
\be
\hspace{-3mm} K^{(2)}U_{12} = \int_{\b_0,\b_{0'}}  \!\!\left(\frac{\z{1}{2}}{\z{1}{0}\z{0}{0'}\z{0'}{2}}\right)^{1{-}\eps}  K^{(2)}_{[1\,00'\,2]}
 \big(U_{10}U_{02}{+}U_{10'}U_{0'2}{-}2U_{10}U_{00'}U_{0'2}\big)
 + \gamma_K^{(2)} K^{(1)}U_{12}\,. \label{NLOk}
\ee
The coefficient $\gamma_K^{(2)}$ can be fixed by matching a certain limit controlled by the cusp anomalous dimension (see section \ref{sec:linear}): $\gamma_K^{(2)}=-\pi^2/3+O(\eps)$. The full two-loop planar evolution is then given as (\ref{NLOk})
which agrees completely with the existing result for the Balitsky-Kovchegov equation \cite{Balitsky:2009xg}.

\subsection{More on virtual corrections}
\label{ssec:virtual}

Although they were not strictly needed to obtain the two-loop result (\ref{NLOk}) (having taken the two-loop cusp anomalous dimension as a known input), it is instructive to explicitly compute the virtual corrections. Learning to handle them expediently will prevent them
from becoming the bane of our existence at higher loops.

Morally, the coefficient $\gamma_K^{(2)}$ is related to the one-loop correction to the single soft current, which has been
obtained long ago (see \cite{Bern:1999ry,Catani:2000pi}):
\ba\hspace{-11mm}
 \frac{F^{(1){\rm bare}}_{[1\,0\,2]}}{F^{(0)}_{[1\,0\,2]}}
&=&  2{\rm Re}\left[ \frac{c_\Gamma}{\eps^2} \left(\frac{e^{-i\pi}\Qsq{1}{0}{2}}{\mu^2}\right)^{-\eps}
 \frac{-\pi\eps}{\sin(\pi\eps)} \right] = \left(\frac{\Qsq{1}{0}{2}}{\mu^2}\right)^{-\eps}\left[
 - \frac{2c_\Gamma}{\eps^2}+\frac{2\pi^2}{3} +O(\eps)\right]\!. \label{one_loop_soft_current}
\ea
Here $c_\Gamma= \frac{\Gamma(1+\eps)\Gamma(1-\eps)^2}{\Gamma(1-2\eps)(4\pi)^{-\eps}}$
is a ubiquitous loop factor. This formula does not depend on the matter content of the theory.
The ``bare" superscript indicates that we have performed ultraviolet renormalization but have not yet subtracted the infrared divergence,
to which we now turn.

Obviously this result is divergent, whereas we're trying
to compute the finite coefficient $\gamma_K^{(2)}$. Of course, what happens as usual is that 
the physics cannot depend on such a ``bare'' quantity but only on renormalized ones.
A useful intuition here is that infrared divergences in bare amplitudes reflect that scattering states are defined in the deep infrared,
and one must always use the renormalization group to evolve the amplitude back to the physical scale $\mu$ of interest,
as detailed in eq.~(\ref{general_renorm}) below.  This will remove all remaining $1/\eps$ poles.
In the present case, the precise renormalization to use, including finite factors, follows from
the other virtual contributions already included in $K$.
First there is the $U_{12}$ term in $K^{(1)}$, predicted above using the KLN theorem, which can multiply the real part of $K^{(1)}$ iterated using relation (\ref{iterationK1}): 
\be
\frac{\s{1}{0}}{\s{1}{v}\s{v}{0}}\theta(\Q{1}{v}{0}<\Q{1}{0}{2})
 +\frac{\s{0}{2}}{\s{0}{v}\s{v}{2}}\theta(\Q{0}{v}{2}<\Q{1}{0}{2})
-\frac{\s{1}{2}}{\s{1}{v}\s{v}{2}}\theta(\Q{1}{v}{2}<\Q{1}{0}{2})\,.
\label{sub1}
\ee
Second there are the $UU$ terms in the two-loop ansatz (\ref{NLOk}):
\be
\frac{\s{1}{0}}{2\s{1}{v}\s{v}{0}} F^{\rm sub}_{[1\,v0\,2]}
 +\frac{\s{0}{2}}{2\s{0}{v}\s{v}{2}}F^{\rm sub}_{[1\,0v\,2]}\,.
\label{sub2}
\ee
It is important to note that both these contributions are expressed in terms of the phase space of two real partons $0$ and $v$,
whereas the one-loop virtual correction to the soft current (\ref{one_loop_soft_current}) is to be integrated over the phase space
of a single parton $\tilde{p}_0$.  We thus have to match these phase spaces somehow.  The crucial requirement is that
the collinear singularities match at the integrand level.  This requires that, in the limit where $0$ and $v$ are collinear,
their \emph{total} energy matches that in the virtual calculation: $a_0+a_v=\tilde{a}_0$.

The simplest way to do this, while respecting Lorentz invariance away from the collinear limit,
is to keep the angles the same, $\tilde{p}_0\propto p_0$, but use $\Q{1}{0}{2}$ and $\Q{1}{v}{2}$
to define Lorentz-covariant energies for the two daughters $0$ and $v$.
Thus we match the above two corrections with the virtual one at total momentum
$\tilde{p}_0 \equiv p_0\frac{\Q{1}{0}{2}+\Q{1}{v}{2}}{\Q{1}{0}{2}}$.
Let us denote as $f^{\rm split}(p_0, p_v)$ the sum over the five terms in (\ref{sub1})-(\ref{sub2}),
or more generally any homogeneous function of $p_0$, $p_v$.
After changing variable from $p_0$ to $\tilde p_0$ the two-parton phase space factorizes as:
\ba
\int_{p_0,p_{v}}\frac{1}{|p_0|^2}f^{\rm split}(p_0, p_v) H^{\rm parent}(\tilde{p}_0) &=& \int_{\tilde{p}_0} \frac{1}{|\tilde{p}_0|^2}\left(\Qsq{1}{\tilde{0}}{2}\right)^{-\eps}H^{\rm parent}(\tilde{p}_0)
\int_{\beta_v}\left(\frac{\z{1}{2}}{\z{1}{v}\z{v}{2}}\right)^{-\eps}
\nonumber\\
&\times&\int_0^1\frac{dx}{[x(1{-}x)]^{1{+}2\eps}}\,f^{\rm split}\left(x\beta_{\tilde{0}}, (1{-}x)C
\beta_v\right)\,,
\label{split_identity}
\ea
with $C=\left(\frac{\z{1}{\tilde{0}}\z{\tilde{0}}{2}}{\z{1}{v}\z{v}{2}}\right)^{1/2}$, $H^{\rm parent}(\tilde{p}_0)$ is an arbitrary test function,
and $x$ and $1{-}x$ represent the (covariant) energy fractions of the two daughters.

The splitting function $f^{\rm split}$ defined by the sum of (\ref{sub1})-(\ref{sub2}) contains complicated angle-dependent step functions,
which come both from the former equation and from those in $F^{\rm sub}$, explicited in eq.~(\ref{F2sub}).
Conveniently, up to a part that is antisymmetric in $x \to 1{-}x$ and therefore cancel upon integration,
all the step functions cancel out except those proportional to $\theta(\Q{1}{v}{2}<\Q{1}{0}{2})$.
Keeping only these surviving terms, and decomposing the sum into two pieces for later convenience, we thus write
$f^{\rm split} \equiv G_{\{1\,\tv0\,2\}} + G_{\{1\,0\tv\,2\}}$ where
\be\begin{aligned}
G_{\{1\,\tv0\,2\}} &\equiv
-\left(\frac{\s{1}{0}}{\s{1}{v}\s{v}{0}}-\frac{\s{1}{2}}{2\s{1}{v}\s{v}{2}}\right)\theta(\Q{1}{v}{2}<\Q{1}{0}{2})
 - \frac{\s{1}{0}}{2\s{1}{v}\s{v}{0}}\left(F_{[1\,v0\,2]} - 1\right),
\\
G_{\{1\,0\tv\,2\}} &\equiv
-\left(\frac{\s{0}{2}}{\s{0}{v}\s{v}{2}}-\frac{\s{1}{2}}{2\s{1}{v}\s{v}{2}}\right)\theta(\Q{1}{v}{2}<\Q{1}{0}{2})
 - \frac{\s{0}{2}}{2\s{0}{v}\s{v}{2}}\left(F_{[1\,0v\,2]} - 1\right)\,.
\end{aligned}
\label{Gdef}
\ee
Stripping off the integral over the radiated gluon momentum $\tilde{p}_0$ in eq.~(\ref{split_identity}),
we then get the total effective soft current:
\be
 F^{(1){\rm ren}}_{[1\,0\,2]}=F^{(1){\rm bare}}_{[1\,0\,2]} + \left(\Qsq{1}{0}{2}\right)^{-\eps}\delta^{(1)} \label{renorm_2loop}
\ee
with $\delta$ the integral over the splitting function:
\ba
\delta^{(1)} &\equiv& 
 -\int_0^1 \frac{dx}{\big[x(1{-}x)\big]^{1{+}2\eps}}\int_{\b_v} \left(\frac{\z{1}{0}}{\z{1}{v}\z{v}{2}}\right)^{-\eps} 
 2\left(G_{\{1\,\tv0\,2\}} + G_{\{1\,0\tv\,2\}}\right)_{p_v=p_0^0\beta_v\frac{(1{-}x)C}{x}}
\nonumber\\
&=& \int_0^{\frac12}\frac{2dx}{[x(1{-}x)]^{1{+}2\eps}}\times
\int_{\b_{v}}\left(\frac{\z{1}{2}}{\z{1}{v}\z{v}{2}}\right)^{{-}\eps}\left(\frac{\z{1}{0}}{\z{1}{v}\z{v}{0}}+\frac{\z{0}{2}}{\z{0}{v}\z{v}{2}}-\frac{\z{1}{2}}{\z{1}{v}\z{v}{2}}\right)
\nl && 
+\int_{\b_{v}} \frac{\z{0}{2}}{\z{0}{v}\z{v}{2}} \left[ 1+ \frac{\z{1}{2}\z{0}{v}}{\z{0}{1}\z{v}{2}-\z{v}{1}\z{0}{2}}\right]\log\frac{\z{0}{1}\z{v}{2}}{\z{v}{1}\z{0}{2}}
=\left(\frac{2c_\Gamma}{\eps^2}-\pi^2\right)+O(\eps). \label{deltaone}
\ea
Note that, although it is defined as a complicated looking integral, $\delta^{(1)}$ is just a constant: this had to be the case
since the integral is manifestly Lorentz-invariant and an homogeneous function of three null vectors, and all such invariants are constant.
Adding it to the bare matrix element (\ref{one_loop_soft_current}) according to (\ref{renorm_2loop}) then gives:
\be
F^{(1){\rm ren}}_{[1\,0\,2]}\equiv \gamma_K^{(2)}  = -\frac{\pi^2}{3} + O(\eps)
\ee
in perfect agreement with the two-loop cusp anomalous dimension recorded below eq.~(\ref{NLOk}).

\section{Three-loop evolution} \label{sec:three-loops}

We now proceed to derive and assemble the ingredients for three-loop infrared divergences.
The chief conceptual issue is to organize the subtraction of subdivergences,
of which there are plenty at three loops.  We would like to (and will) obtain the evolution kernel $K^{(3)}$
as a sum of absolutely convergent integrals involving physical building blocks (the so-called remainder function)
in which we can set $\eps=0$ directly.

\subsection{First ingredient: Triple-soft current}
\label{ssec:triplesoft}

The first building block is the square of the tree-level soft current for emission of three partons.
This needs to be summed over all produced parton species: gluons, fermions or scalars.

The easiest way to obtain it is from the soft limit of the planar four particle integrand, which is amply documented
in the literature. We fix two external legs to be in the matrix element and two in the conjugate,
and sum over all (Cutkoski) unitarity cuts which separate them.  In the relevant limit, where the cut internal propagators become soft, the integrand from the outer square factors out and the outermost cut propagators act as the parent dipole $U_{12}$.

\begin{figure}
        \centering
        \begin{minipage}{0.4\textwidth}
          \centering
          \includegraphics[width=0.8\textwidth]{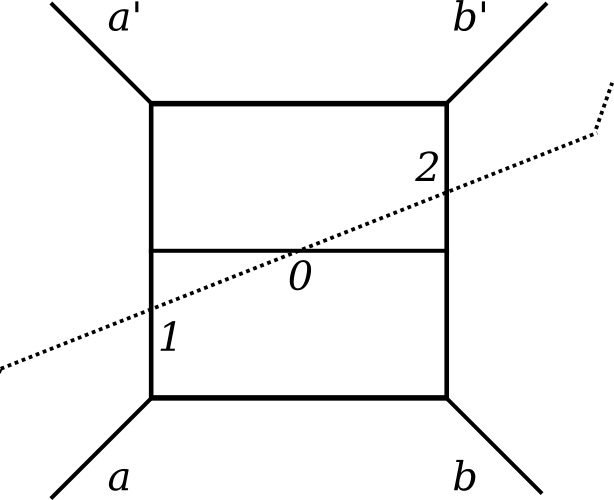}
          \\[5mm]
          (a)
        \end{minipage}%
        \begin{minipage}{0.6\textwidth}
        \centering
        \begin{minipage}{0.5\textwidth}
          \centering
          \includegraphics[width=0.85\textwidth]{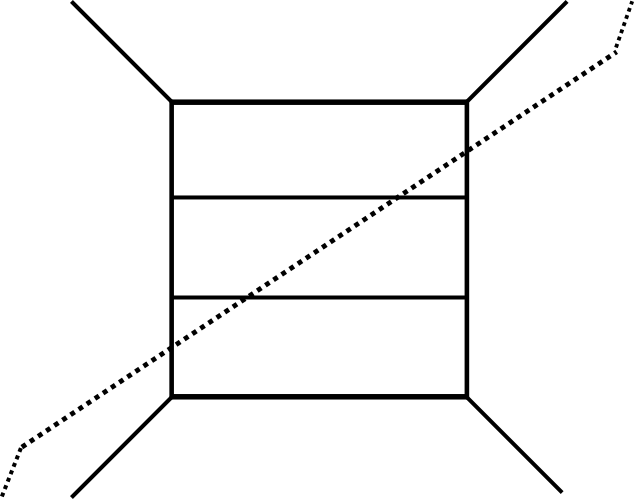}
          \vskip0.7truecm
          \includegraphics[width=0.85\textwidth]{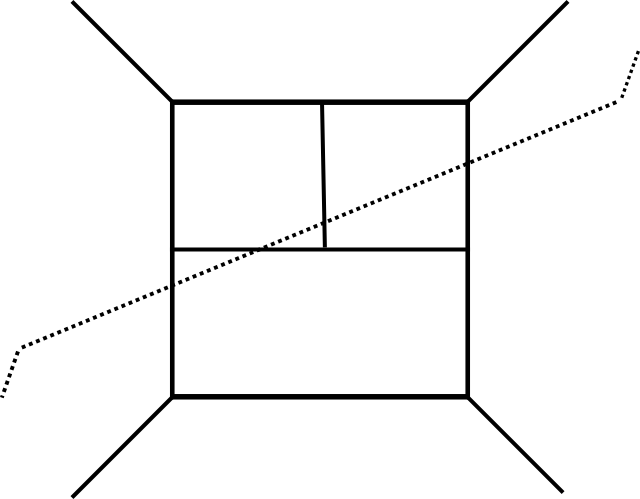}
          \end{minipage}%
        \begin{minipage}{0.5\textwidth}
          \centering
          \includegraphics[width=0.85\textwidth]{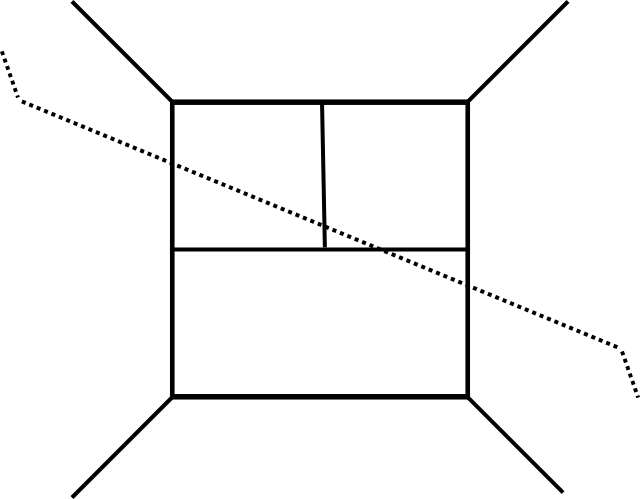}
          \vskip0.7truecm
          \includegraphics[width=0.85\textwidth]{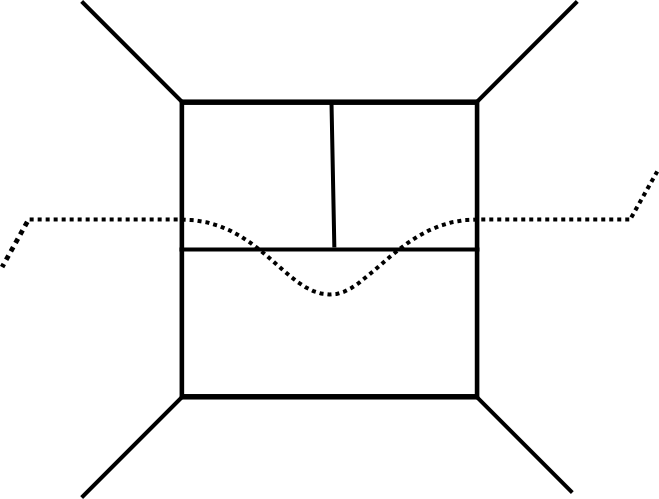}
        \end{minipage}%
          \\[5mm]
          (b)
        \end{minipage}
\caption{Extracting squared soft currents from the four-particle integrand: cuts which give the squares of single (a) and double (b) emissions by taking the cut internal propagators to be soft.}
\label{fig:cuts}
\end{figure}

As an illustration, consider the two-loop integrand, which in planar $\mathcal{N}=4$ is a sum of two double-boxes.
With the momenta labelled as in fig.~\ref{fig:cuts}a:
\be
 \mathcal{I}^{(2)} = \frac{[(p_a+p_b)^2]^2(p_a-p_{a'})^2}{p_1^2p_0^2p_2^2(p_1+p_0)^2(p_2+p_0)^2(p_1-p_a)^2(p_1+p_0-p_{a'})^2} + \mbox{(one permutation)}.
\ee
Taking $p_1$, $p_2$ and $p_0$ to be on-shell with $p_0$ soft, this simplifies to
\be
 p_1^2p_0^2p_2^2\,\mathcal{I}^{(2)}\longrightarrow \frac{(p_a{+}p_b)^2(p_a{-}p_{a'})^2}{(p_1{-}p_a)^2(p_1{-}p_{a'})^2} \times \frac{s_{12}}{s_{10}s_{02}}
\nonumber
\ee
where the first factor is recognized as just the cut of the one-loop amplitude (a scalar box).
Dividing it out leaves the dipole radiator (\ref{F1}),
as expected.  The other three-particle cut of the same diagram (where the cut runs south-east) adds the correct factor 2,
and the rotated double-box is subleading in the soft limit.

Moving on, the three-loop integrand is the
sum of ladders and tennis court scalar integrals (with simple, specific numerators, see \cite{Bern:1997nh}).
Four cuts, shown in fig.~\ref{fig:cuts}b, together with their top-down flips, contribute in the soft limit.
They yield, respectively, the four terms (from left to right and top to down):
\ba
 F_{[1\,00'\,2]} &=& \frac{\s{1}{2}\s{0}{0'}}{2\ssr{1}{00'}\ssl{00'}{2}}+ \frac{\s{1}{0}}{2\ssr{1}{00'}}
 +\frac{\s{0'}{2}}{2\ssl{00'}{2}}+\frac12
 \nl &=&
  1+ \frac{\s{1}{2}\s{0}{0'}+\s{1}{0}\s{0'}{2}-\s{1}{0'}\s{0}{2}}{2\ssr{1}{00'}\ssl{00'}{2}}.
\ea
This is in perfect agreement with the direct calculation recorded in eq.~(\ref{F2}).

Having thus validated the method, it is a simple exercise to extract the square of the triple-soft current from the known 4-loop integrand.
We found the 7-loop package \cite{Bourjaily:2011hi} (recently extended to 8 loops \cite{Bourjaily:2015bpz}) particularly useful for this.
To most usefully record the result, we note that its soft limits are easily predicted.
There are five independent soft limits, where (by factorization) it must reduce to double-soft currents:
\be\begin{aligned}
&F_{[1\,00'0''\,2]}\xrightarrow{0\,\, {\rm soft}} F_{[1\,0'0''\,2]},\quad
F_{[1\,00'0''\,2]}\xrightarrow{0'\,\, {\rm soft}} F_{[1\,00''\,2]},\quad
F_{[1\,00'0''\,2]}\xrightarrow{0''\,\, {\rm soft}} F_{[1\,00'\,2]}, \\
&F_{[1\,00'0''\,2]}\xrightarrow{0\sim0'\,\,{\rm both\,\,soft}} F_{[1\,00'\,0'']},\qquad
F_{[1\,00'0''\,2]}\xrightarrow{0'\sim0''\,\,{\rm both\,\,soft}} F_{[0\,0'0''\,2]}\,. \label{softF3}
\end{aligned}\ee
There are also various double scaling limits, where $F$ reduces to 1.
With a simple ansatz each limit can be accounted for by a single term,
hence leaving a finite remainder:
\ba
 F_{[1\,00'0''\,2]} &=&1+ (1+P) \left( \frac{\s{1}{2}\s{0}{0'}+\s{1}{0}\s{0'}{2}-\s{1}{0'}\s{0}{2}}{2\ssr{1}{00'}\ssl{00'0''}{2}} +
 \frac{\s{1}{0''}\s{0}{0'}+\s{1}{0}\s{0'}{0''}-\s{1}{0'}\s{0}{0''}}{2\ssr{1}{00'}\ss{00'0''}}\right)
 \nl&&+\frac{\s{1}{2}\s{0}{0''}+\s{1}{0}\s{0''}{2}-\s{1}{0''}\s{0}{2}}{2\ssr{1}{00'0''}\ssl{00'0''}{2}}+F^{\rm safe}_{[1\,00'0''\,2]}\,.
\label{F3}
\ea
Here $P$ the parity operation $\{1,0\}\leftrightarrow \{2,0''\}$.
The result we obtain from the four-point integrand matches precisely this form, with the remainder vanishing
in all soft limits. For future convenience we write it here as a sum of individually regular pieces:
\be
F^{\rm safe}_{[1\,00'0''\,2]}=(1+P)(e_1+e_2)+e_3+e_4\,, \label{F3safe}
\ee
\ba
 e_1 &=& \frac{1}{4\ssr{1}{00'0''}\ssl{0'0''}{2}\ss{00'0''}}
\left(\begin{array}{c}\displaystyle
    \s{0}{0''}(2\s{1}{0''}\s{0'}{2}{+}\s{1}{0'}(\s{0'}{2}{-}\s{0''}{2}))
   -\s{0}{0'}(2\s{1}{0'}\s{0''}{2}{+}\s{1}{0''}(\s{0''}{2}{-}\s{0'}{2})) \\\displaystyle
 +\s{0'}{0''}(2\s{1}{0}\s{0''}{2}{-}\ssr{1}{0'0''}\s{0}{2}{-}\s{1}{2}\ssr{0}{0'0''})\end{array}\right),
\nl
 e_2 &=& \frac{\s{1}{0}(\s{1}{2}\s{0'}{0''}+\s{1}{0''}\s{0'}{2}-\s{1}{0'}\s{0''}{2})}{4\ssr{1}{00'}\ssr{1}{00'0''}\ssl{0'0''}{2}},
 \qquad\quad
 e_3 = \frac{\s{1}{0'}\s{0'}{2}}{2\ssr{1}{00'}\ssl{0'0''}{2}} - \frac{\s{1}{0'}\s{0'}{2}+\s{1}{0'}\s{0}{2}+\s{1}{0''}\s{0'}{2}}{2\ssr{1}{00'0''}\ssl{00'0''}{2}},
\nl
 e_4&=& \frac{\s{1}{2}(\s{1}{2}\s{0}{0'}\s{0'}{0''}{+}\s{1}{0}\s{0''}{2}\ss{00'0''})}{4\ssr{1}{00'}\ssr{1}{00'0''}\ssl{00'0''}{2}\ssl{0'0''}{2}}
 + \frac{\s{1}{2}(\s{0}{0'}{+}\s{0'}{0''}{-}\s{0}{0''})}{4\ssr{1}{00'0''}\ssl{00'0''}{2}} -\frac{\s{1}{2}\s{0}{0'}}{4\ssr{1}{00'}\ssl{00'0''}{2}}
 -\frac{\s{1}{2}\s{0'}{0''}}{4\ssr{1}{00'0''}\ssl{00'}{2}}\,.\nonumber
\ea
As a cross-check, we have reproduced numerically the squared soft current (\ref{F3})-(\ref{F3safe}) by a direct
Feynman diagram calculation, summing up the gluon, fermion and scalar contributions,
and also using the computer package \cite{Bourjaily:2010wh}.
For convenience, this formula, and others in this paper, is included in computer-readable format in the ancillary text file \texttt{formulas.txt}, attached to the arXiv submission
of this paper.

\subsection{Second ingredient: Double-soft current and the remainder function}

\def\Log#1{\log\left(#1\right)}
\def\Logsq#1{\log^2\left(#1\right)}

\begin{figure}
        \centering
        \includegraphics[width=0.5\textwidth]{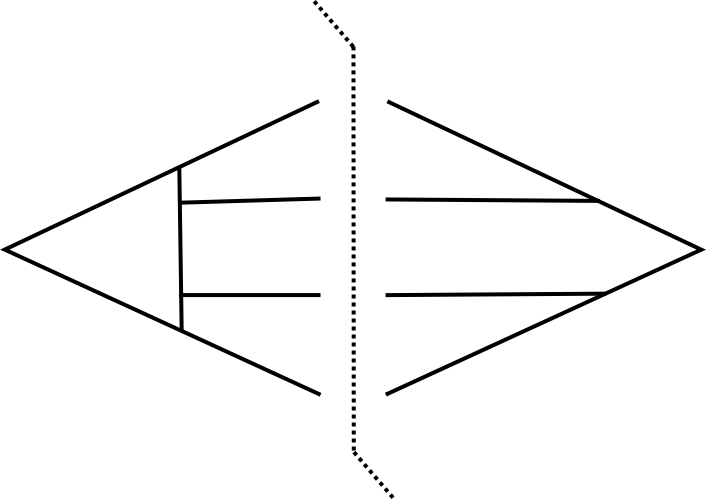}
          \\
\caption{One-loop virtual correction to double soft current contributing to the cross-section at three loops.}
\label{fig:virtual}
\end{figure}

To obtain the one-loop correction to the double soft current in the simplest way, we take the limit of two soft partons
in the known one-loop six-point amplitude. These soft partons can be of any species (gluons, fermions and scalars).
Consider for example the case when the two soft gluons have the same helicity.
In this case we use the one-loop correction to the MHV amplitude (four positive and two negative helicity gluons),
divided by the tree amplitude \cite{Bern:1994zx}:
\ba
 \frac{1}{c_\Gamma}\frac{M_6^{(1){\rm MHV}}}{M_6^{(0){\rm MHV}}} &=&  (1+C^2+C^4)\left[
 \frac{-2}{\eps^2} +\frac{2}{\eps}\Log{\frac{(-\ss{23})(-\ss{56})}{\mu^4}}
+ \Li_2\left(1-\frac{\ss{123}\ss{345}}{\ss{12}\ss{45}}\right) + \frac{\pi^2}{3}
\right.\nl && \hspace{30mm}\left.
-\Log{\frac{(-\ss{23})}{\mu^2}}\Log{\frac{(-\ss{12})(-\ss{34})}{\mu^2(-\ss{56})}}
\right],
\ea
where the operation $C$ is a cyclic rotation by one. 
The one-loop soft current is obtained by taking the limit where partons $2$ and $3$ become the
soft partons $0$ and $0'$, and subtracting the one-loop correction to the parent four-point amplitude.
In this limit, the two color-adjacent partons $1$ and $4$ define the parent dipole, and the other two decouple,
thus giving us the soft current
\ba\!\!\!\!
\frac{1}{c_\Gamma}\frac{\SS^{(1)}_{[1\,0^+0'^+\,2]}}{\SS^{(0)}_{[1\,0^+0'^+\,2]}} &=&
-\frac{2}{\eps^2}+\frac{2}{\eps}\log\frac{-\Qsq{1}{00'}{2}}{\mu^2}
-\Log{\frac{(-\s{1}{0})(-\s{0'}{2})}{\mu^2(-\s{1}{2})}}\Log{\frac{(-\s{0}{0'})}{\mu^2}}
\nl &&
+\Li_2\left(-\frac{\s{1}{0'}}{\s{1}{0}}\right)
+\Li_2\left(-\frac{\s{0}{2}}{\s{0'}{2}}\right)
+\Li_2\left(1-\frac{\ssr{1}{00'}\ssl{00'}{2}}{\s{1}{2}\s{0}{0'}}\right) +O(\eps). \label{softMHV}
\ea
It is important to note that since all invariants are positive (timelike), the Feynman prescription adds an imaginary part
to all logarithms: $\log(-s_{ij})= \log|s_{ij}|-i\pi$.

For soft gluons of opposite helicity, as well as for soft fermions and scalars, one needs the NMHV (super)amplitude
\cite{Bern:1994cg,Drummond:2008vq}.
It may be amusing to note that the two fermions soft current is the same in QCD and $\mathcal{N}=4$ SYM, since the contributing diagrams are the same.
Thus some effective supersymmetry can also be used at one loop in QCD as well.

The component formulas are somewhat involved, and
in the $\mathcal{N}=4$ theory further simplifications occur when summing over particle species in the interference with the tree amplitude.
For this reason, here we record only the final result of the helicity sum, e.g. the one-loop correction
to the squared soft current, in appendix in eq.~(\ref{F2bare}):
\be F^{(1)}_{[1\,00'\,2]} \equiv \left(\frac{4\s{1}{2}}{\s{1}{0}\s{0}{0'}\s{0'}{2}}\right)^{-1}
 \sum_{h_1,h_2} \left(\left[\SS^{(1)}[1\,0^{h_1}0'^{h_2}\,2]\right]^*\left[\SS^{(0)}[1\,0^{h_1}0'^{h_2}\,2]\right]+ {\rm c.c.}\right). \label{defF2bare}
\ee
We used the package in \cite{Bourjaily:2013mma} to cross-check our expressions.

Importantly, as was the case at two loops (and for the MHV example above), this one-loop correction is infrared divergent,
while we expect the physics to depend only on renormalized, finite quantities.
The standard, $\overline{\rm MS}$ way to renormalize is to remove the integral of the infrared anomalous dimension:
\be
 F^{(1){\rm ren},\overline{MS}}_{[1\,00'\,2]}
 \equiv \bar{\mathcal{P}}\exp\left[\int_0^\mu \frac{d\lambda}{\lambda} \gamma_{\rm IR}(\lambda)\right]
 F^{(1){\rm bare}}_{[1\,00'\,2]} \label{general_renorm}
\ee
where at one-loop
$\gamma_{\rm IR}=\frac{g^2_{\rm YM}N_c}{8\pi^2}\log\frac{|s_{10}s_{00'}s_{0'2}|}{|s_{12}|\mu^4}$ for the soft current squared.
This is the conventional definition of so-called hard matrix elements in the SCET literature.
Although a good starting point, this is however not very convenient for us, because we would like to subtract something which has a simple
representation as a phase space integral.

The governing physical principle is that the subtraction should match all singularities of the triple-real emission at the integrand level, in all (single-) soft and collinear limits. This ensures that when we add it back later all divergences will cancel cleanly pre-integration.
Furthermore we would like a simple analytic form for the integrated subtraction.
This can be achieved by defining Lorentz-invariant functions of three angles, like we did in section \ref{ssec:virtual}, since these automatically integrate to constants.

Let us thus consider the general problem of renormalizing an amplitude $F_{[1\,23\ldots\,n]}$ with $(n{-}2)$ soft partons. We want to renormalize it by adding, say at one-loop,
a phase space integral with one additional real parton $v$:
\be
 \int_{p_2,\ldots,p_{n{-}1}} \left(F^{(1){\rm ren}}_{[1\,2\ldots\,n]}-F^{(1){\rm bare}}_{[1\,2\ldots\,n]}\right) \equiv \int_{p_2,\ldots,p_{n{-}1},p_v}
 \Gamma_{[1\,2\ldots\,n]}^v F^{(0)}_{[1\,2\ldots\,n]}\,. \label{renorm}
\ee
There are two constraints.  In the limit where $v$ is soft, the integrand should reduce to minus the square of the soft current, $G_{\{i\,v\,j\}} \equiv -\frac{\s{i}{j}}{\s{i}{v}\s{v}{j}}$,
emitted from all possible regions, minus
that from the parent dipole.
In collinear limits there is a similar factorization,
but with the important distinction that the parent amplitude must be evaluated with the total momentum (here $j=2,\ldots,n{-}1$ and $i=j-1$, $k=j+1$):
\be\begin{aligned}
\Gamma_{[1\,2\ldots\,n]}^v F^{(0)}_{[1\,2\ldots\,n]}
&\xrightarrow{p_v\,\,{\rm soft}} \left(- G_{\{1\,v\,n\}}+\sum_{i=1}^{n-1} G_{\{i\,v\,i{+}1\}} \right)
 F^{(0)}_{[1\,2\ldots\,n]}, \label{soft_collinear_V}\\
 \Gamma_{[1\,2\ldots\,n]}^v F^{(0)}_{[1\,2\ldots\,n]}
&\xrightarrow{p_v{\parallel}p_j} \frac{1}{\s{v}{j}}F_{[i\,jv\,k]}\times F^{(0)}_{[1\,\ldots (p_j{+}p_v)\ldots\,n]}\,.
\end{aligned}\ee
Note that the labels $i$ and $k$ decouple in the collinear limit.
The fact that the argument of the amplitude is shifted to $(p_j{+}p_v)$ is the main complication since it precludes a simple multiplicative solution.
To solve it, recycling the ingredients in the two-loop subtraction (\ref{Gdef}), we define a three-index operator
$\Gamma_{[i\,j\,k]}$ which rescales $p_j$ in whatever it multiplies:
\be
 \Gamma^v_{[i\,j\,k]}F_{[1\,\ldots j\ldots\,n]}\equiv \left(G_{\{i\,\tv j\,k\}}+G_{\{i\,j\tv\,k\}}\right)F_{[1\,\ldots \tilde{j}\ldots\,n]},\qquad
 \tilde{p}_j^\mu \equiv p_j^\mu\left(1+\frac{\Q{i}{v}{k}}{\Q{i}{j}{k}}\right)\,. \label{three_index_operator}
\ee
Then, it is easy to see that all constraints are simultaneously solved by:
\be
 \Gamma^v_{[1\,2\ldots\,n]} \equiv \frac12\sum_{i=2}^{n-1} \left(\Gamma^v_{[1\,i\,i{+}1]}+\Gamma^v_{[i{-}1\,i\,n]}\right).
\ee
Indeed, the three-index $\Gamma^v_{[i\,j\,k]}$ only has collinear singularities in one region $p_v{\parallel}p_j$,
where the spectator labels $i$ and $k$ decouple, so the collinear limits work out.
In the soft limit it approaches $G_{\{i\,v\,j\}}{+}G_{\{j\,v\,k\}}{-}G_{\{i\,v\,k\}}$, and using telescopic
cancellations one can also see that the first of (\ref{soft_collinear_V}) is fulfilled.

Using the change of variable (\ref{split_identity}) and the integral (\ref{deltaone}), the renormalization defined at one-loop by
eq.~(\ref{renorm}) can be rewritten in a more suggestive form:
\be
  F^{{\rm ren}}_{[1\,2\ldots\,n]}= F^{{\rm bare}}_{[1\,2\ldots\,n]}\times
  \Big[\mathcal{V}_{[1\,2\,n]}\mathcal{V}_{[2\,3\,n]}{\cdots} \mathcal{V}_{[n{-}2\,n{-}1\,n]}\Big]\times
  \Big[\mathcal{V}_{[1\,2\,3]}\mathcal{V}_{[1\,3\,4]}{\cdots} \mathcal{V}_{[1\,n{-}1\,n]}\Big],  
\label{many_dipoles_subtraction}
\ee
where
\be
\mathcal{V}_{[i\,j\,k]}=e^{\frac12\delta^{(1)}g^2(\Q{i}{j}{k})} \label{defofV}
\ee
represents the real-emission correction to the soft current squared between legs $i$ and $k$.
We recall that $\delta^{(1)}\approx \frac{2}{\eps^2}$ (\ref{deltaone}) starts with a double pole, and
$g^2(\lambda)\approx(\lambda/\mu)^{-2\eps}g^2(\mu)$ is the $D$-dimensional running coupling.
Physically, the renormalized amplitude is thus obtained by including amplitudes for a sequence of splittings,
each with the coupling evaluated at its natural scale.
(Either of the sequences in the square brackets would work, but we chose to include both and multiply the exponent by $\tfrac12$ for symmetry reasons.)

The renormalized amplitude $F^{\rm ren}$ is finite for any number of points.
At one-loop $F^{(1){\rm ren}}$ is obtained from the bare result (\ref{F2bare}) by the simple substitution given in eq.~(\ref{F2ren_appendix}).
It turns out that $F^{{\rm ren}}$ is closely related to another canonical finite function in $\mathcal{N}=4$ SYM: the Bern-Dixon-Smirnov remainder function \cite{Bern:2005iz}.
This is defined by dividing the amplitude by an ansatz $A^{\rm BDS}_{n{+}2}$
(essentially an exponential of the one-loop MHV amplitude), which makes it finite and dual-conformal invariant and trivializes its collinear limits.
The ansatz has four parameters: three are essentially the constant, order $\eps$ and $\eps^2$ terms in the function called $f(\epsilon)$
in \cite{Bern:2005iz}, which multiplies the one-loop amplitude in the exponent,
while the fourth adds a common multiplicative factor to all $n$-point amplitudes and cancels out for the soft current.
Thus three parameters affect the soft current; comparing with eq.~(\ref{defofV}) it is easy to see that
these three parameters are in one-to-one correspondence with the double-pole, single-pole and constant term in $\delta^{(1)}$,
and that the infrared divergent parts match.
The $\eps^0$ term in our $\delta$ is slightly different because
for five-partons our scheme automatically yields the cusp anomalous dimension $F_{[1\,0\,2]}=\gamma_K$ (see section \ref{sec:linear} for the higher-loop explanation)
whereas by definition the BDS remainder is unity for five-points.  Thus, with a somewhat schematic notation, we can express our renormalized soft current
directly in terms of the BDS remainder to all loop orders:
\be
 F^{{\rm ren}}_{[1\,2\,\ldots\,n]} = (\gamma_K)^{n-2}\times
 \left|R_{n{+}2}\right|^2 \times e^{\gamma_K f_n}, \label{FrenR}
\ee
where $R_{n+2}=A_{n{+}2}/A^{\rm BDS}_{n{+}2}$ is the BDS remainder (e.g. the amplitude $A_{n+2}$ divided by the BDS ansatz $A^{\rm BDS}_{n{+}2}$)
and
$f_n$ an explicitly known (and finite) function equal to the real part of the difference between the 
exponent in eq.~(\ref{many_dipoles_subtraction}) and the squared one-loop MHV soft current (given for $n=4$ in (\ref{softMHV})), which was exponentiated by BDS.  As we will prove shortly, only the $\eps\to0$ limit of the finite $F^{{\rm ren}}$, and thus also BDS remainder, is needed to get the evolution kernel $K$.

Finally, it is interesting to look at eq.~(\ref{FrenR}) in the other direction, going from the soft current to the remainder.
We can count the number of variables on which the soft current $F^{{\rm ren}}$ depends for $m$ soft partons.
Each on-shell parton gives $3m$ degrees of freedom while invariance under two Lorentz generators and one independent rescaling of the $\b_i$ remove 3,
giving $3(m{-}1)$ invariants.
Due to dual conformal symmetry, the $k$-point BDS remainder $A_k/A^{\rm BDS}_{k}$ in eq.~(\ref{FrenR}) depends on $3(k{-}5)$ invariants, which is equal since $k{=}m{+}4$. These two numbers agree!  That is, dual conformal symmetry implies that the soft limit is not lossy,
and we conclude that, through eq.~(\ref{FrenR}), the $\eps\to 0$ limit of BDS remainder in planar $\mathcal{N}=4$ SYM
uniquely determines the renormalized soft current, and \emph{vice-versa}.

\subsection{Nested subtractions for virtual contribution}
\label{ssec:nested_virtual}

Given the renormalized amplitude, it is natural to integrate it over relative energies to obtain a contribution to $K$,
with suitable subtractions as was done in section \ref{ssec:2loop}:
\be
  K^{\rm ren}_{[1\,00'\,2]} \equiv  g^4(\Q{1}{00'}{2})\int_0^\infty\frac{d\tau}{\tau} 2F^{{\rm ren,sub}}_{[1\,(\tau\b_0)\b_{0'}\,2]}\,. \label{K32ren}
\ee
For the subtracted integrand $F^{{\rm ren,sub}}$ it would be tempting to use again eq.~(\ref{F2sub}), but one needs to be more careful
and pay due attention to the renormalization scales of the various objects.
Indeed, as is clear from the renormalization group equation (\ref{integratedRG}), all couplings in the subtractions get evaluated
at their private scales $\Q{i}{j}{k}$,
which are distinct from the common overall scale $\Q{1}{00'}{2}$ that we
assign to $K^{\rm ren}_{[1\,00'\,2]}$.
In addition, the finite parts of the renormalization (\ref{many_dipoles_subtraction}) do not match.
The correct loop-level definition, which accounts for all these effects, is rather
\ba
 F^{\rm ren, sub}_{[1\,00'\,2]} &\equiv& F^{{\rm ren}}_{[1\,00'\,2]}
\nl &&
 -\theta(\Q{1}{0}{0'}{<}\Q{1}{0'}{2})\kappa_{[1\,0\,0'];[1\,0'\,2]}
 F^{{\rm ren}}_{[1\,0\,0']}(g^2(\Q{1}{0}{0'}))F^{{\rm ren}}_{[1\,0'\,2]}(g^2(\Q{1}{0'}{2}))
\nl &&
 -\theta(\Q{0}{0'}{2}{<}\Q{1}{0}{2})\,\kappa_{[0\,0'\,2];[1\,0\,2]} \,F^{{\rm ren}}_{[0\,0'\,2]}(g^2(\Q{0}{0'}{2}))\,F^{{\rm ren}}_{[1\,0\,2]}(g^2(\Q{1}{0}{2})),
\label{Fren_sub}
\ea
where all the couplings are to be evaluated in terms of the common one of the overall process:
$g^2(\lambda)\mapsto g^2(\Q{1}{00'}{2})\left(\frac{\lambda}{\Q{1}{00'}{2}}\right)^{-2\eps}$.
The prefactors, which account for the coupling constants stripped from the two-parton amplitude and for mismatching subtractions
of infrared divergences, are
\ba
\kappa_{[1\,0\,0'];[1\,0'\,2]}&\equiv& \frac{g^2(\Q{1}{0}{0'})g^2(\Q{1}{0'}{2})}{g^4(\Q{1}{00'}{2})}
e^{\frac12\delta^{(1)}\big(g^2(\Q{1}{0}{0'})+g^2(\Q{1}{0'}{2})-g^2(\Q{0}{0'}{2})-g^2(\Q{1}{0}{2})\big)}
\nl 
 &=& e^{g^2\kappa^{(1)}}+O(\eps),\qquad\qquad \kappa^{(1)}= \log\frac{\s{1}{0}\s{0}{2}}{\s{1}{0'}\s{0'}{2}} \log \frac{\s{1}{2}\s{0}{0'}}{\s{1}{0'}\s{0}{2}}\,.
\label{kappa}
\ea
For the other subprocess we get the same but with $-\kappa^{(1)}$.
Specializing to what we need at three loops, extracting the coefficient of $g^2(\Q{1}{00'}{2})$ in $F^{\rm ren,sub}$ and using that
$F^{(1)\rm ren}_{[1\,0\,2]}=-\pi^2/3$ from subsection \ref{ssec:virtual}, this becomes
\be
 F^{(1){\rm ren,sub}}_{[1\,00'\,2]} = F^{(1){\rm ren}}_{[1\,00'\,2]}
 -\theta\big(\Q{0}{0'}{2}{<}\Q{1}{0}{2}\big)\left(\frac{-2\pi^2}{3}+\kappa^{(1)}\right)-\theta\big(\Q{1}{0}{0'}{<}\Q{1}{0'}{2}\big) \left(\frac{-2\pi^2}{3}-\kappa^{(1)}\right).
\label{F1ren_sub}
\ee
The critical conceptual point here is that we won't need the $O(\eps)$ terms in this expression.
This is because the combination in eq.~(\ref{Fren_sub}), in which all objects are defined to all orders in $\eps$,
is precisely the one which vanishes to all order in $\eps$ near the endpoints $\tau\to0$ and $\tau\to\infty$
(this follows from the factorization properties of the bare amplitudes $F^{\rm bare}$).
This precludes any $\eps/\eps$ effect.
The extension to higher loops is clear: one just includes more terms in the expansion of $\delta$. 
Also we expect only minor changes in the presence of a nontrivial $\beta$-function as in full QCD,
where $g^2(\lambda)$ will now be a series in $g^2(\Q{1}{00'}{2})$.

\subsection{Nested subtractions for triple real contribution}

We now turn to the fully real contribution to $K^{(3)}$, which is given by the IR
divergent part of triple-real emission, minus the subdivergences associated with iterations of $K^{(1)}$ and $K^{(2)}$.
The basic idea is to write the subtractions as phase space integrals with step functions, exploiting (\ref{iterationK1}) and its higher-multiplicity generalizations.
In this way all energy sub-divergences (with fixed angles, as appropriate since the angles are fixed by the color rotations $U$)
will cancel under the integration sign. To write the result concisely, we recursively define subtracted integrands $F^{\rm sub}$,
generalizing eq.~(\ref{F2sub}).  Introducing the abbreviations
\be
[X][Y] \equiv F^{\rm sub}_{[X]}F^{\rm sub}_{[Y]}\,\theta(Q^2_{[X]}{<}Q^2_{[Y]}),\qquad
[X][Y][Z] \equiv F^{\rm sub}_{[X]}F^{\rm sub}_{[Y]}F^{\rm sub}_{[Z]}\,\theta(Q^2_{[X]}{<}Q^2_{[Y]}{<}Q^2_{[Z]}),\nonumber
\ee
these are defined as:
\begin{subequations}
\ba
 \Fsuba{1}{0}{2} &\equiv& \Fa{1}{0}{2} = 1,
 \\
 \Fsubb{1}{00'}{2} &\equiv& \Fb{1}{00'}{2} - \br{1}{0}{0'}\br{1}{0'}{2} - \br{0}{0'}{2}\br{1}{0}{2}, \label{Fsub2}
\\
 \Fsubc{1}{00'0''}{2} &\equiv& \Fc{1}{00'0''}{2} -\br{1}{0}{0'}\br{1}{0'0''}{2}-\br{0}{0'}{0''}\br{1}{00''}{2}-\br{0'}{0''}{2}\br{1}{00'}{2}
 \nl && - \br{1}{00'}{0''}\br{1}{0''}{2}-\br{0}{0'0''}{2}\br{1}{0}{2}
\nl &&
-\br{1}{0}{0'}\br{1}{0'}{0''}\br{1}{0''}{2}
-\br{0'}{0''}{2}\br{0}{0'}{2}\br{1}{0}{2}
-\br{0}{0'}{0''}\br{1}{0}{0''}\br{1}{0''}{2}
\nl &&
-\br{0}{0'}{0''}\br{0}{0''}{2}\br{1}{0}{2}
-\br{1}{0}{0'}\br{0'}{0''}{2}\br{1}{0'}{2}
-\br{0'}{0''}{2}\br{1}{0}{0'}\br{1}{0'}{2}. \label{Fsub3}
\ea
\label{Fsub}
\end{subequations}
\!\!The structure is straightforward: there is one subtraction for each possible subprocess (consistent with the planar structure), and the unsubtracted $F$'s are given in eq.~(\ref{F2}) and (\ref{F3}). Intuitively, the $F^{\rm sub}$'s are a device to compute the logarithm of $F$:
the preceding equations can be generated (and generalized to all orders)
by formally solving the equation $\mathcal{P}e^{\int F^{\rm sub}}=\int F$, order by order in the number of emitted partons.

As shown in section \ref{sec:two-loops},  what is relevant for the evolution is the integral over relative energies:
\be
 K^{(3)}_{[1\,00'0''\,2]}\equiv \int_0^\infty \frac{d\tau}{\tau}\frac{d\tau'}{\tau'} 4\Fsubc{1}{(\tau\b_0)(\tau'\b_{0'})\b_{0''}}{2}. \label{K3real}
\ee
Thanks to the pattern of subtractions, and to the factorization of soft currents (see eqs.~(\ref{softF2}) and (\ref{softF3})), 
$\Fsubc{1}{00'0''}{2}$ vanishes in all soft limits and its energy integral at fixed angles is absolutely convergent at all orders in $\eps$.
One might worry that the step functions make it tricky to integrate in practice, but in fact they
always multiply trivial measures like $d\tau/\tau$. Furthermore, the explicit expression (\ref{F3}) naturally splits into
several individually convergent pieces.
For example, the piece $F^{\rm safe}$ doesn't contain any step function and converges by itself.
The pieces from the ``1'' in $\Fa{1}{0}{2}$, $\Fb{1}{00'}{2}$ and $\Fc{1}{00'0''}{2}$ contain multiple step functions,
but all share the trivial measure $d\tau/\tau \,d\tau'/\tau'$ and so immediately integrate to logarithms.
Finally, the five nontrivial subtractions in (\ref{F3}) naturally combine with the remaining terms in (\ref{Fsub3}), to produce five
individually convergent integrals.  

So our problem is reduced to computing finite energy integrals; these produce functions of transcendental weight 2.
A good, systematic way to compute such integrals is the differential equation method described \ref{app:integrals}.\footnote{
For energy integrations the method is considerably simpler than for the transverse integrals illustrated in appendix,
because partial fractions and integration-by-parts in one variable are more elementary and the final contributions are given
from boundary terms instead of contact terms.}
The most difficult integrals are contained within $F^{\rm safe}$.
One of them, in particular, coming from the first line below eq.~(\ref{F3safe}),
cannot be written simply in terms of the angular distances
$\z{i}{j}$, but requires associated spinors ($\b_i^{\alpha\dot{\beta}}\equiv\lambda_i^{\alpha}\tilde{\lambda}^{\dot{\beta}}$):
\ba
f_1&\equiv& \int_0^\infty \frac{d\tau}{\tau} \frac{d\tau'}{\tau'} 4e_1(\tau\b_0,\tau'\b_{0'},\b_{0''})
\nl &=&
2{\rm Re}\left\{\left[1+ \frac{\z{0'}{0''}\l 0\,2\r[2\,1]}{\z{0''}{2}\l 0\,0'\r[0'1]-\z{0'}{2}\l 0\,0''\r[0''1]}\right]
\left[
\Li_2\!\left(1{-}\frac{\z{1}{0'}\z{0''}{2}}{\z{1}{0''}\z{0'}{2}}\right)-\Li_2\!\left(1{-}\frac{\z{0}{0''}\z{0'}{2}}{\z{0}{0'}\z{0''}{2}}\right)+
\right.\right.\nl&&\hspace{0mm}\left.\left.
+ \Li_2\!\left(-\frac{[1\,0][0'\,0'']}{[1\,0''][0\,0']}\right)-\Li_2\!\left(-\frac{\l1\,0\r\l0'\,0''\r}{\l1\,0''\r\l 0\,0'\r}\right)
+\log\frac{\z{1}{0}\z{0'}{0''}}{\z{1}{0''}\z{0}{0'}}\log\frac{\z{0''}{2}\langle 0\,0'\r[0'1]}{\z{0'}{2}\langle 0\,0''\r|[0''1]}
\right]\right\}\!. \label{def_f1}
\ea
Here we have used a commonly used notation for the Lorentz-invariant spinor products:
$\l i\,j\r=\epsilon_{\alpha\beta}\lambda_i^{\alpha}\lambda_j^\beta$ and
$[i\,j]=\epsilon_{\dot\alpha\dot\beta}\tilde\lambda_i^{\dot\alpha}\tilde\lambda_j^{\dot\beta}$ with $\epsilon$ antisymmetric.
(Under the stereographic projection (\ref{stereographic_projection}), these map respectively to: $\l i\,j\r=(z_i{-}z_j)$ and $[i\,j]=(\bar{z}_i{-}\bar{z}_j)$.)
The other integrals are more elementary and produce at most dilogarithms of cross-ratios of $\alpha$'s.

To give the final result we define the five cross-ratios:
\be
u_1 \equiv  \frac{\z{1}{2}\z{0}{0'}}{\z{1}{0'}\z{0}{2}}\,,\quad
u_2 \equiv  \frac{\z{1}{2}\z{0'}{0''}}{\z{1}{0''}\z{0'}{2}}\,,\quad
u_3 \equiv  \frac{\z{1}{2}\z{0}{0''}}{\z{1}{0''}\z{0}{2}}\,,\quad
v_1 \equiv  \frac{\z{1}{0}\z{0'}{2}}{\z{1}{0'}\z{0}{2}}\,,\quad
v_2 \equiv  \frac{\z{1}{0'}\z{0''}{2}}{\z{1}{0''}\z{0'}{2}}\,.\nonumber
\ee
Then the triple-real integral gives
\ba
K^{(3)}_{[1\,00'0''\,2]} &=&  \left(1-\frac{u_3}{1-v_1 v_2}\right)
\left[\begin{array}{l}\displaystyle
2\Li_2\left(1-\frac{1}{v_1 v_2}\right)-2\Li_2\left(1-\frac{1}{v_1}\right)
-2\Li_2\left(1-\frac{1}{v_2}\right)
\vspace{1mm}\\\displaystyle
+\log v_1 \log v_2 +\log(v_1 v_2)\big(\!\log(u_1 u_2) -\tfrac32\log u_3\big)
\end{array}\right]
\nl && +
\left(u_1 u_2 -u_1v_2-u_2v_1 + v_1 + v_2 - u_1 - u_2 + u_3\right)
\left[\Li_2\left(1-\frac{1}{v_1 v_2}\right)-\zeta_2\right]
\nl &&
 + 3\log u_1\log u_2 -\tfrac32\log^2 u_3 + (1+P)(f+f_1), \label{triplereal1}
\ea
where $f_1$ is the special function in eq.~(\ref{def_f1}), $P$ exchanges labels $(1,0)$ and $(2,0'')$ and acts on cross-ratios as $(u_1,v_1){\leftrightarrow}(u_2,v_2)$, and:
\ba
f&=&
\phantom{+}\left(1-\frac{u_1}{1-v_1}\right)\left[2\Li_2\left(1-\frac{1}{v_1}\right)+\log v_1\left(\log\frac{u_2}{v_2}-\frac12\log u_1\right)\right.
\nl &&
\hspace{30mm}\left.+
\big(1+v_2-u_2\big)\left\{\Li_2\left(1-\frac{1}{v_2}\right)-\Li_2\left(1-\frac{1}{v_1 v_2}\right)\right\}\right]
\nl &&
+\left(1-\frac{u_1}{u_3-v_1 u_2}\right)
\left[
\log\frac{v_1 u_2}{u_3} \left(\log\frac{u_2}{v_2}
-\frac32\log\frac{u_1}{u_3} \right)
-2\Li_2\left(1-\frac{v_1 u_2}{u_3}\right)\right]. \label{triplereal2}
\ea

\subsection{Nested subtractions for renormalization counter-terms}

The final ingredient is the ``add'' part from the
 ``add and subtract'' game that led to the renormalized amplitude (\ref{Fren_sub}).
These also have three angular integrations but one fewer color dipole $U$.
In addition there are similar pieces inherited from lower loops,
for example from the term with just $U_{10'}U_{0'2}$ in the two-loop evolution.
It is useful to devise a notation for such terms, like $G_{\{1\,\underline{0}0'\,2\}}$, wherein the underlined index represents the angle from which
a Wilson line is omitted and the curly bracket highlights the presence of a virtual parton.
This is why we've split the virtual correction into two terms ($G_{\{1\,\tv0\,2\}} + G_{\{1\,0\tv\,2\}}$) in eq.~(\ref{Gdef}),
because these two end up with different color structures and so get exponentiated at different scales: $\Q{1}{v0}{2}\neq \Q{1}{0v}{2}$.

Similarly, to exponentiate the three-loop kernel (as would be needed for a putative four-loop calculation), we would
need to specify where $v$ fits within the color structures of $\Gamma^v_{[1\,00'\,2]}$, which determines the relevant scale $Q$. Thus although not strictly necessary here,
it is useful to account for that information because it helps show the internal logic.
Thus we organize the ``add'' terms into three color structures:
\be
K^{(3)}U_{12}\big|^{\rm 3\,angles}= \int_{\b_0,\b_{0'},\b_{0''}}\left[\begin{array}{l}
\phantom{+}K^{(3)}_{[1\,00'0''\,2]}\frac{\z{1}{2}}{\z{1}{0}\z{0}{0'}\z{0'}{0''}\z{0''}{2}}(-2U_{10}U_{00'}U_{0'0''}U_{0''2})
\\
+K^{(3)\rm{add}}_{\{1\,\underline{0}0'0''\,2\}}\frac{\z{1}{2}}{\z{1}{0'}\z{0'}{0''}\z{0''}{2}}(-2U_{10'}U_{0'0''}U_{0''2})
\\
+K^{(3)\rm{add}}_{\{1\,0\underline{0}'0''\,2\}}\frac{\z{1}{2}}{\z{1}{0}\z{0}{0''}\z{0''}{2}}(-2U_{10}U_{00''}U_{0''2})
\\
+K^{(3)\rm{add}}_{\{1\,00'\underline{0}''\,2\}}\frac{\z{1}{2}}{\z{1}{0}\z{0}{0'}\z{0'}{2}}(-2U_{10}U_{00'}U_{0'2})
\end{array}\right].\label{K3add}
\ee
Here we only show the terms in $K^{(3)}$ with three angular integrations, two has been dealt with
in subsection \ref{ssec:nested_virtual} and one will be dealt with shortly.
The underlined index shows the variable whose Wilson line and radiator factor are omitted.
The angular functions are the integrals over relative energies of corresponding $G^{\rm sub}$'s,
\be
 K^{(3)\rm{add}}_{\{1\,\underline{0}0'0''\,2\}} = \int_0^\infty \frac{d\tau}{\tau}\frac{d\tau'}{\tau'} G^{\rm sub}_{\{1\,\underline{(\tau \beta_0)}(\tau'\beta_{0'})\beta_{0''}\,2\}}\,,\quad \mbox{etc.}
\ee
The $G^{\rm sub}$'s contain two ingredients. First, there is the difference between the renormalized $F^{\rm ren,sub}$ in eq.~(\ref{Fren_sub})
and the corresponding bare expression:
\be\begin{aligned}
 G^v_{[1\,00'\,2]} \equiv \Gamma_{[1\,00'\,2]}^v F^{\rm sub}_{[1\,00'\,2]}
 &+(\Gamma^v_{[1\,0\,0']}+\Gamma^v_{[1\,0'\,2]})\theta(\Q{1}{0}{0'}{<}\Q{1}{0'}{2})
\\ &+(\Gamma^v_{[0\,0'\,2]}+\Gamma^v_{[1\,0\,2]})\theta(\Q{0}{0'}{2}{<}\Q{1}{0}{2}).
\end{aligned}\ee
Second, there is the subtraction of everything inherited from virtual corrections at lower loops:
from the $U_{12}$ term at one-loop (\ref{K1}), and from the $(U_{10}U_{02}+U_{10'}U_{0'2})$ part of two-loop (\ref{NLOk}).
To allow their subsequent exponentiation, the result is to decomposed into 3 color structures:
\be
G^{\rm sub}_{\{1\,\tv 00'\,2\}}+G^{\rm sub}_{\{1\,0\tv 0'\,2\}}+G^{\rm sub}_{\{1\,00'\tv\,2\}}\equiv G^v_{[1\,00'\,2]} -\mbox{lower subtractions}\,.
\ee
A simple systematic color decomposition for the subtractions can be done as follows.
Whenever, in a subprocess, both indices adjacent to $v$ are the same as in the considered $G^{\rm sub}$,
we weight this contribution by 1; when only one index is shared, we weight by $\frac12$, and when none is shared, we weight by 0.
For example, consider the following term coming from the real part of $K^{(1)}$ times the virtual part of $K^{(2)}$:
\be
 F^{\rm sub}_{[1\,0\,0']}G^{\rm sub}_{\{1\,\tv0'\,2\}}\theta(\Q{1}{0}{0'}{<}\Q{1}{\tv0'}{2}).
\ee
We place half of this term into $G^{\rm sub}_{\{1\,\tv00'\,2\}}$ and half into $G^{\rm sub}_{\{1\,0\tv0'\,2\}}$, because
$v$ occurs between $1$ and $0'$.
Using these rules to generate the subtractions recursively, using the same notation as in eq.~(\ref{Fsub3})
(writing $\{a\,b\ldots\, c\}\equiv G^{\rm sub}_{\{a\,b\ldots\, d\}}$ and inserting a step function between each bracket, either curly or square),
then gives
\begin{subequations}
\ba
 G^{\rm sub}_{\{1\,\tv\,2\}} &\equiv& G_{\{1\,\tv\,2\}} = -\frac{\s{1}{2}}{\s{1}{v}\s{v}{2}}
\\
 G^{\rm sub}_{\{1\,\tv0\,2\}} &\equiv& \gm{1}{\tv0}{2}\br{1}{0}{2}-\big(\brv{1}{\tv}{0}-\tfrac12\brv{1}{\tv}{2}\big)\br{1}{0}{2},
\label{Gsub_2loopa}
\\
  G^{\rm sub}_{\{1\,0\tv\,2\}} &\equiv& \gm{1}{0\tv}{2}\br{1}{0}{2}-\big(\brv{0}{\tv}{2}-\tfrac12\brv{1}{\tv}{2}\big)\br{1}{0}{2},
\label{Gsub_2loopb}
\\
 G^{\rm sub}_{\{1\,\tv00'\,2\}}&\equiv&
 \gm{1}{\tv00'}{2}\br{1}{00'}{2}
 +\big(\gm{1}{\tv0}{0'}+\tfrac12\gm{1}{\tv0'}{2}\big)\br{1}{0}{0'}\br{1}{0'}{2}
  +\gm{1}{\tv0}{2}\br{0}{0'}{2}\br{1}{0}{2}
\nl &&
 -\tfrac12\br{1}{0}{0'}\brv{1}{\tv0'}{2}-\brv{1}{\tv0}{0'}\br{1}{0'}{2}-\br{0}{0'}{2}\brv{1}{\tv0}{2}
\nl &&
 -\big(\brv{1}{\tv}{0}-\tfrac12\brv{1}{\tv}{2}\big)\big(\br{1}{00'}{2}+\br{1}{0}{0'}\br{1}{0'}{2}+\br{0}{0'}{2}\br{1}{0}{2}\big)
\nl && -\br{1}{0}{0'}\big(\tfrac12\brv{1}{\tv}{0'}-\tfrac12\brv{1}{\tv}{2}\big)\br{1}{0'}{2}
-\br{0}{0'}{2}\big(\brv{1}{\tv}{0}-\tfrac12\brv{1}{\tv}{2}\big)\br{1}{0}{2}\,,
\label{Gsub_3loopa}
\\
 G^{\rm sub}_{\{1\,0\tv0'\,2\}}&\equiv&
 \gm{1}{0\tv0'}{2}\br{1}{00'}{2}
 \nl &&
  + \big(\gm{1}{0\tv}{0'}+\tfrac12\gm{1}{\tv0'}{2}\big)\br{1}{0}{0'}\br{1}{0'}{2}
 + \big(\gm{0}{\tv0'}{2}+\tfrac12\gm{1}{0\tv}{2}\big)\br{0}{0'}{2}\br{1}{0}{2}
 \nl &&
-\tfrac12\br{1}{0}{0'}\brv{1}{\tv0'}{2}-\brv{1}{0\tv}{0'}\br{1}{0'}{2}-\tfrac12\br{0}{0'}{2}\brv{1}{0\tv}{2}
 -\brv{0}{\tv0'}{2}\br{1}{0}{2}
\nl &&
-\brv{0}{\tv}{0'}\big(\br{1}{00'}{2}+\br{1}{0}{0'}\br{1}{0'}{2}+\br{0}{0'}{2}\br{1}{0}{2}\big)
\nl &&-\br{1}{0}{0'}\tfrac12\brv{1}{\tv}{0'} \br{1}{0'}{2}
-\br{0}{0'}{2}\tfrac12\brv{0}{\tv}{2} \br{1}{0}{2}\,,
\label{Gsub_3loopb}
\\
 G^{\rm sub}_{\{1\,00'\tv\,2\}}&\equiv& P  G^{\rm sub}_{\{2\,\tv0'0\,1\}},
\label{Gsub_3loopc}
\ea \label{Gsub}
\end{subequations}
\!\!where $P$ is the parity $(10){\leftrightarrow}(0'2)$.
This looks messy, but the upshot is that the internal logic is straightforward and
the terms can be automatically generated to any desired order.
(Formally, the terms can be generated by series-expanding the schematic formula
$\mathcal{P}e^{\int (F^{\rm sub}+G^{\rm sub})}= \int (F+G)e^{\int K|_{U_{12}}}$.\footnote{
The fully virtual correction $e^{\int K|_{U_{12}}}$ to the parent dipole $U_{12}$ appears on the right-hand side since the soft currents $F$
are defined to act on the bare amplitude; this is also the reason why $\{1\,\tv\,2\}$ appears with opposite sign wherever it does.})
This generalizes the subtractions used at two loops: the energy integral of $G^{\rm sub}_{\{1\,\tv0\,2\}}$ matches the $U_{10'}U_{0'2}$ term in eq.~(\ref{NLOk}).
Although we haven't defined the individual $\gm{1}{\tv00'}{2}$ (only their sum $\Gamma^v_{[1\,00'\,2]}$) we expect that a definition exists which will make the
energy integrals converge absolutely for each of the color structure $G^{\rm sub}_{\{1\,\tv00'\,2\}}$, as this is certainly the case for the sum
which is all we need here at three loops.  Furthermore, by construction,
the collinear singularities of $K^{(3){\rm add}}$ cancel exactly those of $K^{(3)}$, to all orders in $\eps$,
so it is apparent that $O(\eps)$ corrections to any kernel are not needed.

Thus we only need to compute the finite integral (\ref{K3add}) with $\eps=0$. The integrated result turns out to be
somewhat inelegant, so we decided to replace it by a simpler counter-term with the same collinear singularity.
From inspection of the triple-real result (\ref{triplereal1}), we find divergences as $0{\parallel}0'$ or $0'{\parallel}0''$,
and also in the double scaling limit $0{\parallel}0'{\parallel}0''$, but not  when one or two partons become collinear to $1$ or $2$.
A simple counter-term which removes the divergence as $0'{\parallel}0''$ is:
\be
 K^{(3)c.t.}_{[1\,00'0''\,2]} = 
\left[\left(1+\frac{\z{1}{2}\z{0}{0'}}{\z{1}{0}\z{0'}{2}{-}\z{1}{0'}\z{0}{2}}\right)\log \frac{\z{1}{0}\z{0'}{2}}{\z{1}{0'}\z{0}{2}}
+\frac32\log{\frac{\z{1}{2}\z{0}{0'}}{\z{1}{0'}\z{0}{2}}}\right]\log{\frac{\z{1}{2}\z{0}{2}\zsq{0'}{0''}}{\z{1}{0''}\z{0}{0''}\zsq{0'}{2}}}. \label{K3ct}
\ee
To construct an integral that is also absolutely convergent in double collinear limits,
we can easily play with the color structures, exploiting that $U_{ij}\to 1$ when $i{\parallel}j$.
Arranging for each color factor to separately fulfill the KLN theorem (vanishing when $U_{ij}=1$),
the full three-loop evolution is then written as (\ref{K3final}) below, 
where the difference compared to subsection \ref{ssec:nested_virtual} is simply:
\be
 K^{(3)}_{[1\,00'\,2]} - K^{(3)\rm{ren}}_{[1\,00'\,2]} = \int_{\b_v} \left[ (1+P)\frac{\z{0'}{2}}{\z{0'}{v}\z{v}{2}}K^{(3)c.t.}_{[1\,00'v\,2]}
 +K^{(3)\rm{add}}_{\{1\,\tv00'\,2\}}+K^{(3)\rm{add}}_{\{1\,0\tv0'\,2\}}+K^{(3)\rm{add}}_{\{1\,00'\tv\,2\}}\right] \nonumber
\ee
with $P$ the symmetry $(10){\leftrightarrow}(20')$.
This is again an absolutely convergent integral which can be done at $\eps=0$,
using the methods of appendix \ref{app:integrals}.
We find a surprisingly compact result:
\be
 -\left(1+\frac{\z{1}{2}\z{0}{0'}}{\z{1}{0}\z{0'}{2}{-}\z{1}{0'}\z{0}{2}}\right)\left[\log^2\left(\frac{\z{1}{2}\z{0}{0'}}{\z{1}{0'}\z{0}{2}}\right)+4\zeta_2\right]\log\left(\frac{\z{1}{0}\z{0'}{2}}{\z{1}{0'}\z{0}{2}}\right)
 -\frac{11}{6}\log^3\left(\frac{\z{1}{2}\z{0}{0'}}{\z{1}{0'}\z{0}{2}}\right). \label{subtraction_change}
\ee
Its simplicity (compared with eqs.~(\ref{Gsub})) suggests that an even simpler organization of the subtractions could exist.
Adding this result to the energy integral of the one-loop remainder function (\ref{F1ren_sub},\ref{F2bare},\ref{F2ren_appendix}),
computing using the same method explained above, we thus obtain the part of the evolution with two angular integrals.

\subsection{Final result: The three-loop BK equation in planar $\mathcal{N}=4$ SYM}\label{ssec:final}

In summary, we have computed the three-loop correction to the Balitsky-Kovchegov rapidity evolution equation
(or equivalently Banfi-Marchesini-Smye equation for non-global logarithms) in planar $\mathcal{N}=4$ SYM,
in terms of absolutely convergent integrals over squared amplitudes (or BDS remainder).  The subtraction of subdivergences
has been organized around the physical principle of factorization (see eqs.~(\ref{Fren_sub},\ref{Fsub},\ref{Gsub})),
in such a way that all cancellations are manifest at the integrand level and valid to all orders in $\eps$. This allowed us
to set $\eps=0$ directly in all integrals and be completely certain that we did not miss any $\eps/\eps$ effect.

We have used the squared amplitude for triple-real emission and also the one-loop correction
to double-real emission (related to the one-loop six-point remainder function). In addition $K$ receives contribution
from single-real emission at two-loops, and fully virtual corrections. However, it is not necessary to explicitly compute them.
As mentioned already, fully virtual corrections follow simply from the KLN theorem. And by Lorentz symmetry (kept manifest at all stages of our calculation)
the single-real emissions can only produce a constant $\gamma_K^{(3)}$
time one-loop.  As argued (and tested) in the next section, provided that the $U_{12}$ color structure
appears nowhere else in our expression, what multiplies one-loop must be the cusp anomalous dimension
(known to all loops \cite{Beisert:2006ez}):
$\gamma_K\equiv \tfrac14\Gamma_{\rm cusp}=\frac{g^2_{\rm YM}N_c}{16\pi^2}\left(1-\frac{\pi^2}{3}\frac{g^2_{\rm YM}N_c}{16\pi^2}+\frac{11\pi^4}{45}\left(\frac{g^2_{\rm YM}N_c}{16\pi^2}\right)^2+\ldots\right)$.

Thus our final result for the three-loop BK equation, recalling the lower loop results, is:
\begin{subequations}
\ba 
K^{(1)}U_{12} &=& \int_{\beta_0} \frac{\z{1}{2}}{\z{1}{0}\z{0}{2}}\big(2U_{12}-2U_{10}U_{02}\big), \label{K1final}\\
K^{(2)}U_{12} &=& -\frac{\pi^2}{3} K^{(1)}U_{12} + \int_{\b_0,\b_{0'}} \frac{\z{1}{2}}{\z{1}{0}\z{0}{0'}\z{0'}{2}} K^{(2)}_{[1\,00'\,2]}
 \big(U_{10}U_{02}{+}U_{10'}U_{0'2}{-}2U_{10}U_{00'}U_{0'2}\big), \nl\label{K2final} \\
K^{(3)}U_{12} &=& \frac{11\pi^4}{45} K^{(1)}U_{12} + \int_{\b_0,\b_{0'}} \frac{\z{1}{2}}{\z{1}{0}\z{0}{0'}\z{0'}{2}} K^{(3)}_{[1\,00'\,2]}
 \big(U_{10}U_{02}{+}U_{10'}U_{0'2}{-}2U_{10}U_{00'}U_{0'2}\big)
 \nl && +
 \int_{\b_0,\b_{0'},\b_{0''}}
\frac{\z{1}{2}}{\z{1}{0}\z{0}{0'}\z{0'}{0''}\z{0''}{2}}
\left[K^{(3)}_{[1\,00'0''\,2]}\big(2U_{10'}U_{0'2}{-}2U_{10}U_{00'}U_{0'0''}U_{0''2}\big)\right.
\nl &&\hspace{35mm}\left.
 -(1+P) \left(K^{(3)c.t.}_{[1\,00'0''\,2]} \big(2U_{10'}U_{0'2}-2U_{10}U_{00'}U_{0'2}\big)\right)\right]\!,
 \label{K3final}
\ea
\label{Kfinal}
\end{subequations}
\!\!where $P$ is the parity $(10){\leftrightarrow}(20'')$, $\z{i}{j}\equiv |z_i{-}z_j|^2$ are transverse distances
and $\int_{\beta_0}\equiv\int \frac{d^2z_0}{\pi}$. (Equivalently, for the non-global-logarithmic problem,
the stereographic projection (\ref{stereographic_projection}) gives $\z{i}{j}\equiv\frac{1{-}\cos\theta_{ij}}{2}$
and $\int_{\beta_0}\equiv \int \frac{d^2\Omega_0}{4\pi}$).

The two-loop transverse function $K^{(2)}_{[1\,00'\,2]}$ was given in eq.~(\ref{K2real}), and the triple-real function
$K^{(3)}_{[1\,00'0''\,2]}$ and counter-term $K^{(3)c.t.}_{[1\,00'0''\,2]}$ are in eqs.~(\ref{triplereal1}) and (\ref{K3ct}).
Finally, defining cross-ratios $u$ and $v$ and associated complex numbers $x,\bar{x}$,
\be
u \equiv x\xb =\frac{\z{1}{2}\z{0}{0'}}{\z{1}{0'}\z{0}{2}}\,,\qquad
v \equiv (1-x)(1-\xb) = \frac{\z{1}{0}\z{0'}{2}}{\z{1}{0'}\z{0}{2}}\,, \label{crossratiosuv}
\ee
the effective single-virtual kernel (the sum of eqs.~(\ref{K32ren}) and (\ref{subtraction_change}))  is given as
\ba
 K^{(3)}_{[1\,00'\,2]} &=&
\left(1-\frac{u}{1-v}\right)\log v\left[\log u\log \frac{v}{u}-\frac13\log^2v-4\zeta_2\right]
  +2(1+v-u)\left(\zeta_2\log\frac{u}{v}-2\zeta_3\right) 
\nl && +
   \left(\frac{2u}{1-v}+v-u-1\right)\left[4\Li_3\left(1-\frac{1}{v}\right)+2\Li_2\left(1-\frac{1}{v}\right)\log\frac{v}{u}\right]-\frac56\log^3u
 \nl &&+ 4\big(\Li_3(x)+\Li_3(\xb)-2\zeta_3\big) -2\big(\Li_2(x)+\Li_2(\xb)+2\zeta_2\big)\log u\,. \label{K3virt}
\ea
For convenience, these formulas are reproduced in computer-readable format in the ancillary text file \texttt{formulas.txt}, attached to the arXiv submission of this paper.

We note that eq.~(\ref{K3virt}) is a \emph{single-valued} combination of polylogarithms.
That is, it does not have any branch cut for physical angles (where $x$ and $\bar{x}$ are complex conjugate of each other: $\bar{x}=x^*$,
as is easily verified).
This has to be the case since the kernel represents a physical probability for radiation and there can't be multiple answers for a given set of angles.
Concretely, although this is not manifest, one can verify that the series expansion of the last line around $x=\xb=1$
contains only single-valued logarithms of the type $\log(1-x)(1-\xb)$, but $\log(1-x)$ never appears separately from $\log(1-\xb)$.

\section{Linearized evolution and BFKL Pomeron trajectory}
\label{sec:linear}

In many applications to the high-energy limit, especially those involving dilute targets and projectiles,
the Wilson lines remain close to unity and the physics is governed by a linearized version of eqs.~(\ref{Kfinal}).
Then we set
\be
U_{ij} =1-\UU_{ij}
\ee
and treat $\UU_{ij}$ as a small quantity.
Generically, in the `t Hooft large $N_c$ limit, $\UU_{ij}\sim 1/N_c^2$ when scattering objects made of a fixed number of partons,
or for example a four-point correlator of single-trace operators.
The resulting linear equation is referred to as the BFKL equation and its eigenvalue $j=1-K$ is the Pomeron Regge trajectory.
With this application in mind, in this section we will use the language of transverse plane and conformal symmetry,
instead of the stereographically equivalent language of angles and Lorentz symmetry.

Linearizing the color structures in the three loop result (\ref{K3final}) produces many terms, but these turn out to organize simply into the combination
which appears already at two loops:
\be
 U_{10}U_{02}+U_{10'}U_{0'2}-2U_{10}U_{00'}U_{0'2} \quad\mapsto\quad
 \UU_{10'}+\UU_{02}-\UU_{10} -\UU_{0'2} -2\UU_{00'}\,. \label{linear5terms}
\ee
This is due to an exact symmetry: the large $N_c$ theory is invariant under
the local gauge transformations $U_{ij}\to U_{ij}e^{\alpha_i- \alpha_j}$, representing
independent $U(1)$ gauge transformations in the past and future.
(Beyond the planar limit, only the global SU$(N_c)_{\rm past}\times$SU$(N_c)_{\rm future}$ survives as a symmetry of the Balitsky-JIMWLK equation.)
The combination (\ref{linear5terms}) is the only one invariant under the linear transformation
$\UU_{ij}\mapsto \UU_{ij}+\alpha_i-\alpha_j$, which does not contain
$\UU_{12}$ and is invariant under the parity $(10){\leftrightarrow}(20')$.
That parity is automatic for any conformally-invariant function of four transverse points $1,0,0',2$, and so not really an assumption.

The first four terms on the right of eq.~(\ref{linear5terms}) naively integrate to zero,
\be
 \int\! d^2z_0 d^2z_{0'} (U_{10}-U_{10'}) \frac{\z{1}{2}\,K_{[1\,00'\,2]}}{\z{1}{0}\z{0}{0'}\z{0'}{2}}
\stackrel{{\rm naively}}{=}
\int\!d^2z_0\frac{\z{1}{2}\,U_{10}}{\z{1}{0}\z{0}{2}}
 \int \!d^2z_{0'} \left[\frac{\z{0}{2}\,K_{[1\,00'\,2]}}{\z{0}{0'}\z{0'}{2}}-(1{\leftrightarrow}2)\right]  =0, \label{naive_zero}
\ee
because by conformal symmetry the $z_{0'}$ integral can only produce a constant and thus cannot be antisymmetric in $1$ and $2$.
In the first equality we have used the just-mentioned parity symmetry to trade $(0{\leftrightarrow}0')$ for $(1{\leftrightarrow}2)$.
A subtlety however is that in this rewriting the middle integral fails to be absolutely convergent
in the double scaling limit $z_0,z_{0'}\to z_{2}$, even though the left-hand side is. Due to this, the conformal symmetry argument breaks down
for $z_0=z_2$, enabling a contact term $\delta^2(z_0{-}z_2)$ to appear. (See appendix E of ref.~\cite{Balitsky:2009xg} for explicit examples.)

Taking into account the possibility of such contact terms by adding a constant $C^{(L)}$,
the linearization at $L$-loops takes the general form
\be
 K^{(L)} \UU_{12} = \left(\gamma_K^{(L)}K^{(1)} + 2C^{(L)}\right) \UU_{12}+\int \frac{d^2z_0d^2z_{0'}}{\pi^2} \frac{(-2)\z{1}{2}\,\UU_{00'}}{\z{1}{0}\z{0}{0'}\z{0'}{2}}K^{(L){\rm lin}}_{[1\,00'\,2]}\,, \label{linear2}
\ee
where at two loops $K^{(2){\rm lin}}_{[1\,00'\,2]}=K^{(2)}_{[1\,00'\,2]}$ and at three loops
\be
 K^{(3){\rm lin}}_{[1\,00'\,2]} = K^{(3)}_{[1\,00'\,2]}+2\int \frac{d^2z_v}{\pi} \frac{\z{0'}{2}}{\z{0'}{v}\z{v}{2}} \left(K^{(3)}_{[1\,00'v\,2]}-K^{(3)c.t.}_{[1\,00'v\,2]}\right). \label{3looplinear}
\ee
This integral, like others in this paper, is absolutely convergent.
The factor of two accounts for the contribution with $0$ and $0''$ interchanged, which produces the same result
due to the parity symmetry.  We computed this integral using the differential equation method explained in appendix \ref{app:integrals}.
The resulting function of the cross-ratios $x,\bar{x}$ (defined in eq.~(\ref{crossratiosuv}))
has 5 letters in its symbol: $x,\bar{x},1{-}x,1{-}\bar{x},1-v$, where $v=(1{-}x)(1{-}\bar{x})$. At transcendental weight 3 there exists rather few such functions 
that are real and single-valued in the physical region $\bar{x}=x^*$, in the sense explained below (\ref{K3virt}).
We have found only three nontrivial ones $O_{1,2,3}$.
Since there is limited information content in these functions themselves, we record them in appendix in eq.~(\ref{app:O123})
and here record the concise coordinate space expression for the BFKL kernel ($u=x\bar{x}$, $v=(1{-}x)(1{-}\bar{x})$):
\ba
 K^{(3){\rm lin}}_{[1\,00'\,2]} &=&
 2\left(1-\frac{u}{1-v}\right)\left[ 6O_1+3O_2+6\Li_3(1-v)-2\Li_3(1-v^{-1})-\log(u^2v)\Li_2(1-v)
 \right.\nl&&\hspace{30mm}\left.+\tfrac12\log^3v-\log u\log^2v-\tfrac32\log^2u\log v+3\zeta_2\log v+24\zeta_3\right]
 \nl &&
 +(1+v-u)\left[ -3O_1-3O_2+6\Li_3(1-v^{-1})-2\Li_3(1-v)-\log(u^2 v^{-1})\Li_2(1-v^{-1})
 \right.\nl&&\hspace{30mm}\left.-\tfrac56\log^3v+\tfrac32\log u\log^2v+8\zeta_2\log u-9\zeta_2\log v-30\zeta_3\right]
 \nl && -3(x-\xb)O_3-\tfrac43\log^3u+8O_1\,.
\label{3looplinear_final}
\ea
\begin{figure}
        \centering
        \includegraphics[width=0.7\textwidth]{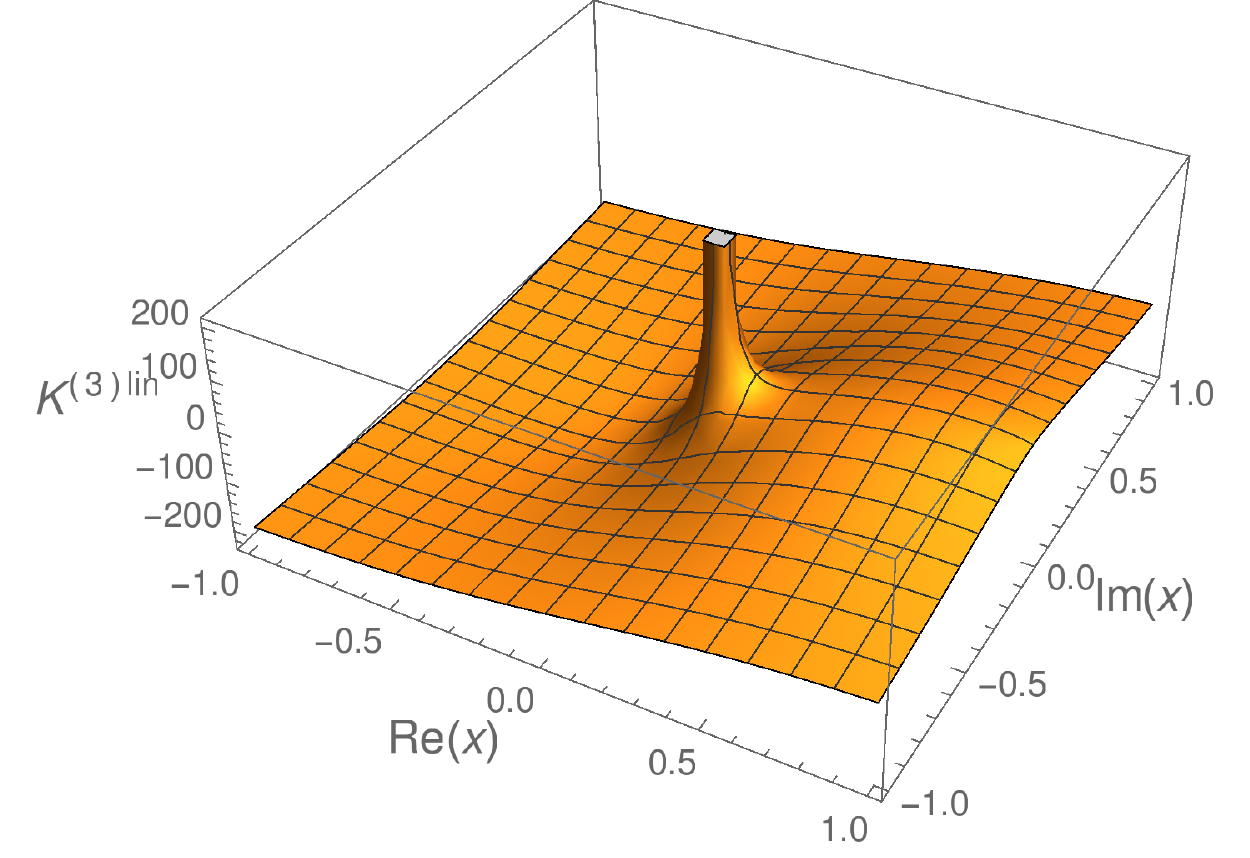}
          \\
\caption{The linear kernel $K^{(3){\rm lin}}$ in coordinate space in the physical region $\bar{x} = x^*$.}
\label{fig:plot_K3lin}
\end{figure}

Now the linear equation (\ref{linear2}) can be diagonalized explicitly because its eigenfunctions are determined by conformal symmetry
\cite{Lipatov:1985uk}. The eigenvalue depends on two quantum numbers: a scaling dimension $\nu$ and an (integer) angular momentum $m$.
It can be extracted by looking at the translation invariant wavefunctions\footnote{In conventional Regge theory, trajectories $j(t)$ are functions of the transverse momentum squared.
In a conformal theory there is a continuum of such trajectories for each value of $t$, but this continuum
is generated by a symmetry and with fixed $p$ and $\nu$ the spectrum becomes discrete, see for example \cite{Brower:2006ea}.} 
\be
\UU_{ij}\equiv \UU(z_i-z_j)=|z_i{-}z_j|^{1+i\nu}[(z_i-z_j)/(\bar{z}_i-\bar{z}_j)]^{m/2}.  \label{wavefunction}
\ee
A special eigenfunction is $\UU(z_i{-}z_j)=z_i-z_j$, corresponding to $m=1$ and $\nu=0$, which is
a generator of the aforementioned U(1) gauge symmetry at large $N_c$.
The eigenvalue of $K=1-j$ must thus vanish in this case, which leads to an exact prediction
for the intercept of the Odderon trajectory at large $N_c$ \cite{Kovchegov:2012rz,Caron-Huot:2013fea}:
\be
 j(m{=}1,\nu{=}0) =1. \label{odderon_intercept}
\ee
Since this property was manifest in the original starting point, {\it i.e.} eq.~(\ref{linear5terms}) before eq.~(\ref{naive_zero}) was used,
in practice we will use this property to fix the constant $C^{(L)}$.
Translation-invariance of the trial wavefunctions enables one transverse integral to be done explicitly.
This simplifies the evolution (\ref{linear2}) to:
\be
K^{(L)}\UU(x)=\left(\gamma_K^{(L)}K^{(1)}+2C^{(L)}\right)\UU(x)-\int \frac{d^2y}{|y|^2}\,H^{(L)}(y)\,\UU(xy)\,,
\ee
where, labelling four points as $\{z_1,z_0,z_{0'},z_2\}=\{1,z,z{-}y,0\}$, the translation-invariant kernel is
\be
H^{(L)}(y)=\int \frac{d^2z\,2K^{(L){\rm lin}}_{[1\,z\,(z-y)\,0]}}{\pi |1-z|^2|y-z|^2}. \label{translation_invariant_case}
\ee
Plugging in the wavefunction (\ref{wavefunction}), we see that the Pomeron trajectory is the Mellin transform of $H^{(L)}(y)$.

The parity symmetry of $K^{(L){\rm lin}}$ makes $|y| H^{(L)}(y)$ invariant under the inversion $y\to 1/y$.
The eigenvalue can thus be written as the sum of two terms,
analytic in the lower- and upper-half $\nu$-planes respectively, representing the contributions from $|y|<1$ and $|y|>1$.
Following a common notation in the literature, we thus write the Regge trajectory $j=1-K$ as:
\be
j(m,\nu)=1+ \sum_{L=1}^\infty \left(\frac{g^2_{\rm YM}N_c}{16\pi^2}\right)^L \left(F^{(L)}_{m,\nu}+F^{(L)}_{m,-\nu}\right)
\label{j_from_F}
\ee
where, for $L>1$,
\be
F^{(L)}_{m,\nu} = \gamma_K^{(L)}F^{(1)}_{m,\nu} -C^{(L)} +\int_{|y|<1} \frac{d^2y}{|y|^2} |y|^{1{+}i\nu}(y/\bar{y})^{m/2}\,H^{(L)}(y)\,. \label{Mellin}
\ee
In summary, the Pomeron trajectory $j(m,\nu)$, defined conceptually as the eigenvalue of the evolution
(\ref{linear2}) on the eigenfunctions (\ref{wavefunction}),
is obtained at three loops
by computing the Mellin transform (\ref{Mellin}) of the translation invariant projection (\ref{translation_invariant_case}) of
the coordinate space kernel (\ref{3looplinear_final}).

\subsection{Result for the eigenvalue}

To efficiently integrate (\ref{translation_invariant_case}) we used the differential equation method, wherein derivatives are
iteratively computed and simplified using integration-by-parts identities.
This method has a long and successful history in the context of dimensionally regulated Feynman integrals \cite{Kotikov:1990kg,Gehrmann:1999as}.
We used a variant that exploits simplifications occuring for absolutely convergent integrals, based on ideas in refs.~\cite{Henn:2013pwa,Caron-Huot:2014lda}.
Our procedure is illustrated in appendix \ref{app:integrals} in a few examples.
The result is an expression for $H^{(3)}(y)$ in terms of iterated integrals starting from the origin $y=0$.

In principle these iterated integrals could be rewritten in terms of polylogarithms,
but we found this neither illuminating nor useful in practice.
Rather, to extract the eigenvalue, we found it more efficient to perform
the angular integration directly at the level of the iterated integral, using again the differential equation method to obtain
the result as a iterated integral in the radial variable $x=|y|^2$.
The radial functions then turned out to be conventional harmonic polylogarithms.
At one- and two-loop this procedure gives expressions that are very uniform for all transverse angular momentum
$m$ ($m\geq 0$):
\be\begin{aligned}
 F^{(1)}_{m,\nu} &= 4\int_0^1 \frac{dx}{x}x^{\frac{1{+}|m|{+}i\nu}{2}}\frac{1}{(1-x)_+} = 4\left[\psi(1)-\psi\left(\frac{1+m+i\nu}{2}\right)\right],
\\
 F^{(2)}_{m,\nu} &= \frac{-\pi^2}{3}F^{(1)}_{m,\nu} +12\zeta_3
 +8\!\int_0^1 \frac{dx}{x}x^{\frac{1{+}m{+}i\nu}{2}} \left[\frac{H_{0,0}}{x-1} +\frac{(1+x^{-m})H_2+(1-(-x)^{-m})H_{-1,0}}{x+1}\right]_{\rm reg.}
\label{newF2eigen}
\end{aligned}\ee
Here the `reg.' notation is an instruction to subtract all the negative powers of $x$ (and powers of $\log(x)$ they multiply)
from the series expansion of the bracket around $x=0$.
The harmonic polylogarithms (with omitted argument $x$) are defined recursively
as \cite{Remiddi:1999ew,Maitre:2005uu}\footnote{Using classical functions: $H_2(x)=\Li_2(x)$, $H_{-1,0}(x)=\Li_2(-x)+\log(x)\log(1{+}x)$ and $H_{0,0}(x)=\frac12\log^2(x)$.}
\be
 H_{\pm i,a_2\ldots,a_n}\!(x) = \int_0^x \frac{dx'}{1\mp x}\frac{\log^{i{-}1}(x/x')}{(i-1)!} H_{a_2\ldots,a_n}\!(x'),
 \qquad H_{0,\ldots,0}(x)=\frac{\log^k x}{k!}\,.
\ee
The concise expression (\ref{newF2eigen})  for the two-loop eigenvalue is apparently new, but we have verified that it agrees, for all values of $|m|$,
with the known result in $\mathcal{N}=4$ SYM \cite{Fadin:1998py,Ciafaloni:1998gs,Kotikov:2002ab}.

Equations (\ref{newF2eigen}) takes the form of a Mellin transform over harmonic polylogarithms, which is well-known to give
harmonic sums (see appendix \ref{app:eigen}), which in the case of eq.~(\ref{newF2eigen}) would have argument
$\frac{1+|m|+i\nu}{2}$ and $\frac{1-|m|+i\nu}{2}$.  However, it is important to note the ``reg'' subscript in that equation, which implies that a number of powers
of $x$ terms, which grows with $|m|$, have to be subtracted.  It would be interesting to see if the result can be usefully written as some kind
of ``regulated'' harmonic sum.

\begin{figure}
        \centering
        \includegraphics[width=0.7\textwidth]{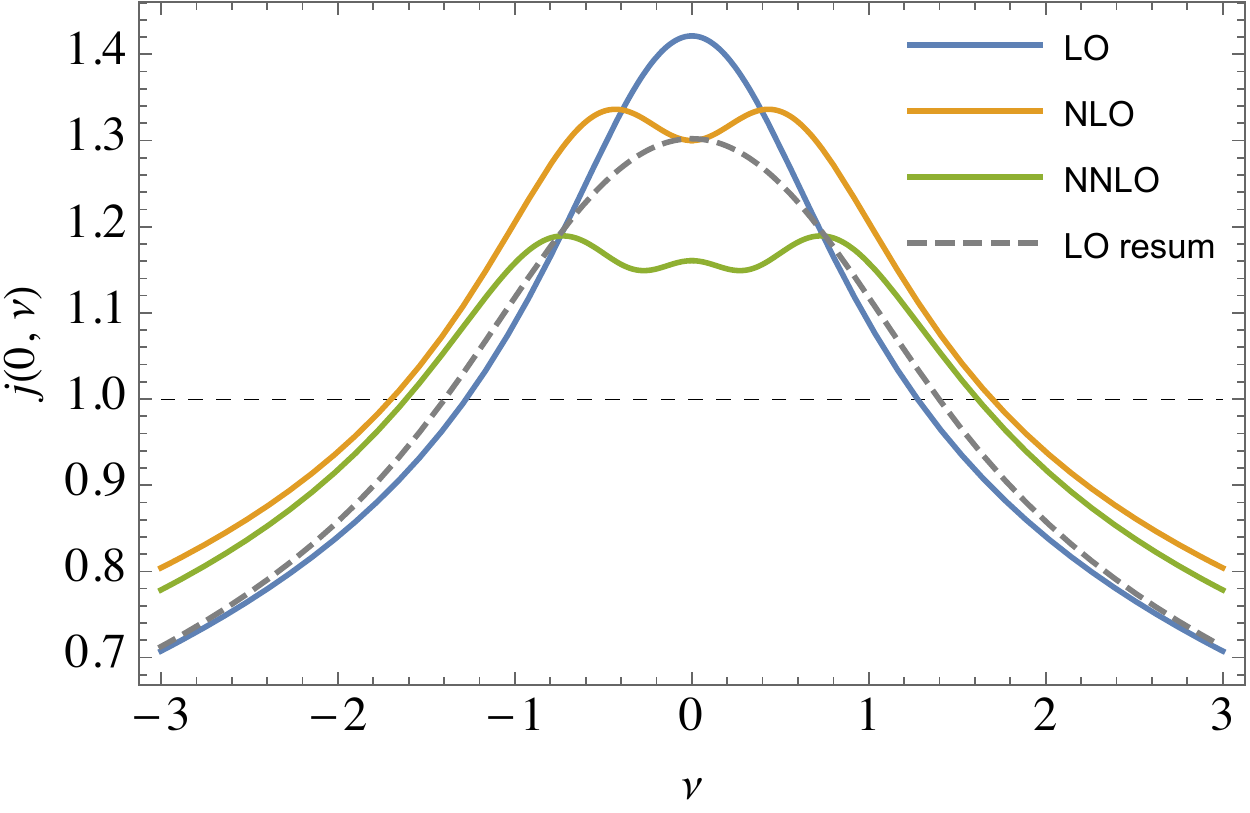}
          \\
\caption{The BFKL eigenvalue for $m=0$ along the real $\nu$ axis at various orders for
 $\lambda = g_{\rm YM}^2 N_c = 6$. Convergence near the maximum is visibly slower than away from it.
  The ``resummation of leading-order'' is defined below eq.~(\ref{level_crossing}). }
\label{fig:plot_m0}
\end{figure}

At three-loops, although the ``reg.'' notation seems to help, we did not succeed in finding a compact formula accounting for the full $m$ dependence, and so
here we restrict our attention to individual values of $m$, for example
\ba
 F^{(3)}_{0,\nu} &=& \frac{11\pi^4}{45}F^{(1)}_{0,\nu} -16(\zeta_2\zeta_3+5\zeta_5)+
 32\int_0^1 \frac{dx}{x}x^{\frac{1{+}i\nu}{2}} \left[ \frac{H_{0,0,0,0}}{1-x} +\frac{f_0^{(3)}}{1+x}\right], \label{F0Mellin}
 \\
f_0^{(3)}&=&-2 H_4 - 2H_{-3, 0} - 4 H_{-1, 3} + H_{3, 0} + 4 H_{3, 1} + 2 H_{-2, -1, 0} - H_{-2, 0, 0} + 2 H_{-1, -2, 0} -2 H_{-1, 2, 0}
\nl && - 8 H_{-1, 2, 1} - 2 H_{2, 0, 0} + \zeta_2(H_{-2}+H_{-1,0}-2H_2-2H_{0,0}) +3 \zeta_3(\tfrac32 H_0-H_{-1})-10\zeta_4.\nonumber
\ea

The Mellin integral in eq.~(\ref{F0Mellin}) gives a practical and efficient way to compute the eigenvalue numerically
for any desired value of $\nu$.  The result of the integral can also be formally expressed in terms of harmonic sums
(see eq.~(\ref{F0mellin1})), although evaluating these sums for complex $\nu$ then requires an analytic continuation.
In appendix \ref{app:eigen} we also provide harmonic sums expressions for $m=1$.

Interestingly, the same constant $C^{(3)}=16(\zeta_2\zeta_3+5\zeta_5)$, fixed here analytically from the condition (\ref{odderon_intercept}),
also appears in the large-spin limit of twist-two anomalous dimensions ($\Delta-2-j\to 8\gamma_K(\log(j)+\gamma_E)-C$),
and in the large-$\nu$ limit of the color-adjoint BFKL kernel \cite{Basso:2014pla}.

A Mathematica notebook \texttt{trajectories\_3loop.nb}
attached to the arXiv submission article allows to evaluate the eigenvalues for any $m$ and $\nu$.
(The command \texttt{j3Eval[m,nu]} evaluates numerically to high accuracy the 3-loop correction to $j(\texttt{m},\texttt{nu})$,
by numerically integrating the series-expansion around 0 and 1 of the radial functions;
the command \texttt{F3integrandHPL[m]} produces symbolic expressions for the radial function and transverse spin $m$
in terms of harmonic polylogarithms.)

For even $m=2,4,6\ldots$, something new happens: the integrand requires a generalization of
harmonic polylogarithms involving iterations of $\int \frac{d}{dx'}\log\frac{1{-}i\sqrt{x'}}{1{+}i\sqrt{x'}}$.
This is related to the square-root containing entries of the symbol of $H(y)$ recorded at the end of appendix \ref{app:integrals}.
While still straightforward to evaluate the Mellin transform numerically, the result cannot be written in terms of
conventional harmonic sums and it is an interesting open problem to characterize this new class of sums.

\begin{figure}
        \centering
        \includegraphics[width=0.7\textwidth]{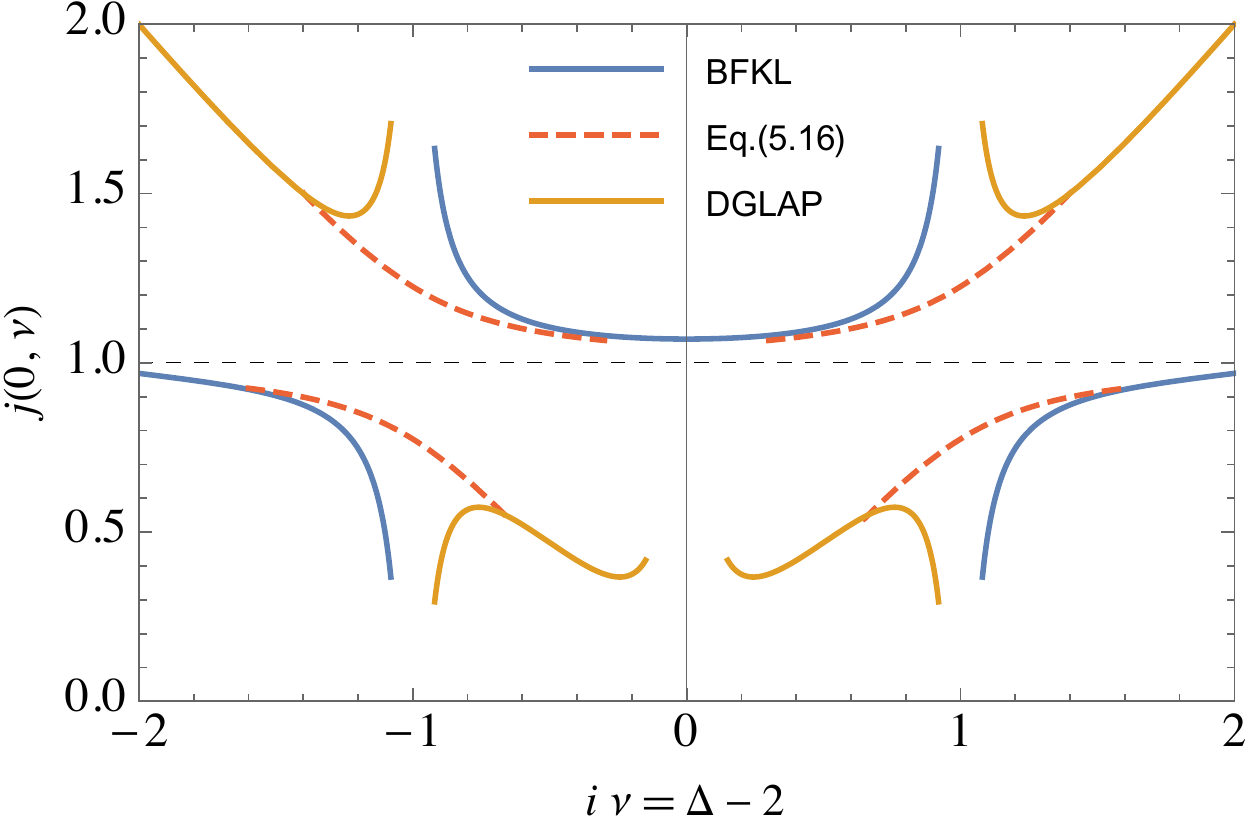}
          \\
\caption{Level repulsion between the Pomeron and DGLAP trajectories for $m=0$ as a function of scaling dimension,
illustrating the $\nu=\pm i$ singularities. (LO expressions plotted with $\lambda = g_{\rm YM}^2 N_c = 1$.)}
\label{fig:plot_level_crossing}
\end{figure}

Finally, we have compared our result for $m=0$ with the recent works \cite{Gromov:2015vua,Velizhanin:2015xsa},
which exploited, respectively, integrability of the theory and high-loop data in the collinear limit.
After converting to our basis, we found perfect agreement with both references (showing in particular that they agree
with each other). The coordinate space kernel (\ref{3looplinear_final}), its corresponding eigenvalue for $m> 0$, and
the nonlinear terms in eq.~(\ref{K3final}), are new predictions.

\pagebreak
\subsection{Collinear singularities and resummation}

\begin{figure}
        \centering
        \includegraphics[width=0.7\textwidth]{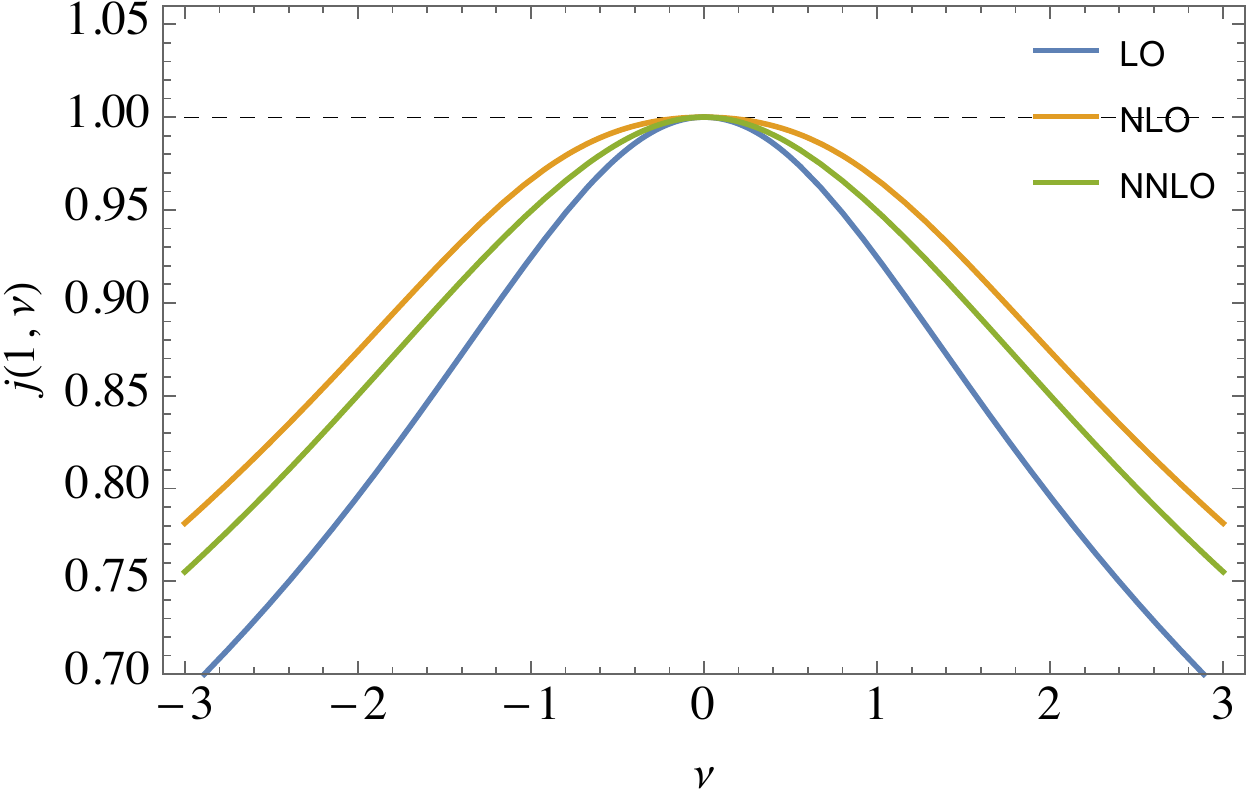}
          \\
\caption{The BFKL eigenvalue for $m=1$ along the real $\nu$ axis at various orders for
 $\lambda = g_{\rm YM}^2 N_c = 6$.}
\label{fig:plot_m1}
\end{figure}

The eigenvalue is plotted for $m=0$ and $m=1$ in figs.~\ref{fig:plot_m0}-\ref{fig:plot_m1}. It is apparent that, especially near the peak for $m=0$,
the perturbative series suffers from slow convergence. This was observed already at two loops and
explained in terms of nearby singularities in the complex plane at $i\nu=\pm1$ \cite{Salam:1998tj}.

In short, these singularities are related to the collinear limit of BFKL,
where the scaling dimension $\Delta=2+i\nu=3$ of the exchanged state coincides with that of twist-two operators: $\Delta=2+j+\gamma(j)$
with $j$ close to 1, e.g. the operators entering the DGLAP equation. As is common for two-level quantum systems,
this crossing of two energy levels \cite{Brower:2006ea} gets resolved 
as depicted in fig.~\ref{fig:plot_level_crossing}:
\be
 j\approx 1+\frac{\Delta-3 \pm \sqrt{(\Delta-3)^2+32g^2}}{2},\qquad \Delta=2+i\nu. \label{level_crossing}
\ee
At small $g^2\equiv \frac{g^2_{\rm YM}N_c}{16\pi^2}$,
one branch choice gives the near-horizontal BFKL trajectory while the other gives the 45${}^\circ$ twist-two (DGLAP)
trajectory.  (The square root formula follows easily by solving $\Delta(j)\approx j+2+\frac{8g^2}{j-1}$  for the $j$,
within the overlapping regime of validity of BFKL and DGLAP
$g^2\ll |j{-}1|\ll 1$ where the anomalous dimension $\gamma(j)$ can be approximated by its leading pole.)
It was shown that, expanding the square root to order $g^4$,
reduces by half the magnitude of the two-loop corrections to the intercept $j(0,0)$ (if one also includes the complex conjugate singularity at $i\nu=-1$) \cite{Salam:1998tj}.  The ``LO resummation'' curve in fig.~\ref{fig:plot_m0}, called ``scheme 2'' in ref.~\cite{Salam:1998tj}, 
thus shows the LO trajectory plus eq.~(\ref{level_crossing}) minus its $O(g^2)$ expansion.
(It would be useful to develop a NLO resummation and we leave it as an open problem for the future.)

The formula (\ref{level_crossing}), expanded to three loops, turns out to
not predict very well the three-loop correction to the intercept
$j(0,0)\approx 1+11.09g^2-84.08g^4-2543.05g^6+O(g^8)$.
In fact it gets even the sign wrong.  By looking at the singular terms in $F$ close to the pole we can try to understand why:
\be
 F_{0,\nu}\xrightarrow{i\nu{\to}1} \frac{8g^2}{\delta} -\frac{64g^4}{\delta^3}
 + g^6\left(\frac{1024}{\delta^5}-\frac{512\zeta_2}{\delta^3} -\frac{576\zeta_3}{\delta^2}-\frac{464\zeta_4}{\delta}\right) +\mbox{regular}
+O(g^8), \label{singular_terms_3loops}
\ee
where $\delta= 1-i\nu$.
Comparing with eq.~(\ref{level_crossing}), we find that the leading pole $1024g^6/\delta^5$ is exactly as predicted (as it had to).
Setting $\delta=1$, the subleading poles however also give a numerically large contribution to the intercept $2F$,
so truncating to the leading pole does not give a good approximation to the intercept.
However, summing up all the singular terms in eq.~(\ref{singular_terms_3loops}),
one finds that about $80\%$ of the three-loop correction to the intercept is reproduced.
A heuristic explanation is that the contributions from the next singularities, at $i\nu=\pm 3$, are suppressed by their distance.

Interestingly, all polar terms at $L$-loops can be obtained from the $L$-loop DGLAP equation.
(See for example \cite{Marzani:2007gk,Costa:2012cb}.) From the higher-loop DGLAP equation
one can get nonsingular terms in the expansion (\ref{singular_terms_3loops}), see for example eq.~(21) of \cite{Velizhanin:2015xsa}.
We have verified that our result (\ref{F0Mellin}-\ref{F0mellin1}) agrees with all these constraints.\footnote{
Compared to eq.~(21) of \cite{Velizhanin:2015xsa} (version 1), we have $\omega\mapsto -\omega$,
to match with the generally accepted convention $\omega=j-1$ that we are following.}

We conclude that the physical picture of \cite{Salam:1998tj}, that large corrections to the intercept originate from
the $i\nu=\pm 1$ collinear singularities, is consistent with the three-loop trajectory we obtained, although the full polar part,
predicted by DGLAP (as opposed to just the leading pole), must be retained.
In general it would be very interesting to find a way to make full use of the DGLAP information at a given loop order

Finally, we comment on the Mellin transform of the level-crossing formula (\ref{level_crossing})
back to coordinate space. The transform produces a Bessel function:
\be
 \int_{-\infty}^{+\infty} \frac{d\nu}{2\pi} |z|^{i\nu-1} \sqrt{(i\nu-1)^2+32g^2}=
 32g^2 \frac{J_1(4g\sqrt{2}\log|z|)}{4g\sqrt{2}\log|z|}\,.
\ee
The right-hand side has appeared in
coordinate space and momentum space resummations \cite{Vera:2005jt,Iancu:2015vea}, so it is nice to see how it
arises form the familiar two-level crossing formula (\ref{level_crossing}).

\section{Discussion and conclusion}\label{sec:conclusion}

In this paper we have computed, for the first time, the evolution equation
which resums large rapidity logarithms in forward scattering to three loops in a gauge theory.
Our main results are the full nonlinear equation (\ref{Kfinal}) in planar supersymmetric Yang-Mills theory,
its linearization (\ref{3looplinear_final}), characterizing the BFKL Pomeron in impact parameter space,
as well as its eigenvalue, the Pomeron Regge trajectory, described in appendix \ref{app:eigen}.
This result is a first step toward the analogous QCD result, and by itself can already be used to
assess the convergence of perturbation theory and its proposed resummations,
and shed light on nonlinear saturation effects at finite coupling.

This computation was made possible thanks to a recently established correspondence with
the resummation of large so-called non-global logarithms, which occur when soft radiation is excluded from a fixed angular region.
This correspondence is helpful because it makes available a body of knowledge on the factorization of infrared and collinear divergences,
and at a conceptual level it defines in a clear way the evolution equation to all loop orders.
This allowed us to derive a systematic subtraction method for nested subdivergences, embodied in eq.~(\ref{Fsub}), such that
all energy integrals at fixed angle become convergent.  We then dealt with collinear subdivergences and real-virtual cancellations
in a second step, by multiplying and dividing by the corrections to the single soft current as in eq.~(\ref{many_dipoles_subtraction}).

Therefore, although we set up our calculation in dimensional regularization and some divergent intermediate objects appeared,
we find that in the end the evolution equation depends only on the $\eps\to0$ limit of physical scheme-independent
quantities like the the Bern-Dixon-Smirnov remainder (\ref{FrenR})! This opens a new possibility to relate an object with the topology of the cylinder, the BFKL Pomeron, to
the integrable system appearing in planar scattering amplitudes \cite{Basso:2015uxa}; graphically speaking, this cuts the cylinder into two half-pipes.

As a highly nontrivial test, of both our calculation and of the integrability approach, we have compared our extracted Pomeron trajectory (\ref{F0Mellin})
with the recent predictions for $m=0$ in \cite{Gromov:2015vua,Velizhanin:2015xsa}, and found perfect agreement!
We have also found perfect agreement, in the collinear limit, with the prediction from anomalous dimensions of twist-two operators.
The trajectory for other transverse angular momenta $m$, and nonlinear interactions, are new predictions which it would be very interesting to check within
the integrability approach.

It is important to clarify the $1/N_c$ counting in which our result is valid.
The projectile is assumed to be made of a finite $\sim N_c^0$
number of Wilson lines, but whose expectation values across the target
can be finite, $\frac{1}{N_c}{\rm Tr}[U_1 U_2^\dagger]\sim 1$.
This asymmetric setup, motivated for example in proton-nucleus collisions,
is the same as that for which the Balitsky-Kovchegov equation is strictly derived.
In the context of AdS/CFT this counting would apply to e.g. a light probe of a black hole. This is also a well-defined setup
and in fact it would be interesting to work out the nonlinear terms in the Balitsky-Kovchegov equation at strong coupling $\lambda$,
including perhaps $1/\sqrt{\lambda}$ stringy effects.
The linear terms, which govern correlators of light operators with large but not-to-large energies (before
the onset of saturation, such that $1{-}U\sim s^{j_0-1}/N_c^2\ll1$) have already been identified with graviton exchange \cite{Brower:2006ea}.

There are several directions in which this work could be extended.
One is to go beyond the planar limit at weak coupling, where the two-loop corrections have recently become available
\cite{Balitsky:2013fea,Kovner:2013ona,Caron-Huot:2015bja}.
Interesting new physical effects appear at three loops in the non-planar sector,
for example the $4{\to}2$ reggeon transition which ``closes the Pomeron loop''
and restores the symmetry between the target and projectile would first be seen there (see for example \cite{Caron-Huot:2013fea}).
Through the KLN theorem, the three-loop evolution could also independently predict from real corrections,
and thus test, the recent result for three-loop soft anomalous dimension \cite{Almelid:2015jia}.
Another direction is towards QCD: technically, our setup gives direct access
to the evolution equation for non-global logarithms, which in QCD will differ from rapidity evolution
by terms proportional to the $\beta$-function.
These could thus be calculated subsequently by calculating matter loops on both sides of the correspondence.

\acknowledgments{
SCH would like to thank Kolya Gromov and Vitaly Velizhanin for discussions regarding the integrability prediction for the trajectory.
SCH's work was supported partly by the Danish FNU Grant No.~126152
and by the Danish National Research Foundation (DNRF91).
MH is supported by the Villum Foundation Grant No.~YIP/VKR022599.}

\begin{appendix}

\section{One-loop correction to the squared double soft current}
\label{app:sixpoint}

Here we record the interference of the tree and one-loop double soft current, defined in eq.~(\ref{defF2bare}),
obtained from the soft limit of the six point amplitude as explained in the text.
\ba
 2F^{(1){\rm bare}}_{[1\,00'\,2]} &=&
 \left(\frac{\s{1}{0}\s{0}{2}-\s{1}{0'}\s{0'}{2}}{\ssr{1}{00'}\ssl{00'}{2}-\s{1}{2}\s{0}{0'}}\right)
  \Log{\frac{\s{1}{0}\ssl{00'}{2}}{\ssr{1}{00'}\s{0'}{2}}}\Log{\frac{\s{1}{2}\s{0}{0'}}{\ssr{1}{00'}\ssl{00'}{2}}} \nl
 && + (1+P)\left(\frac{\s{1}{2}\s{0}{0'}-\s{1}{0'}\s{0}{2}-\s{1}{0}\s{0'}{2}}{\s{1}{0'}\ssl{00'}{2}}\right)
 \Log{\frac{\s{1}{0}}{\ssr{1}{00'}}}\Log{\frac{\s{1}{2}\s{0}{0'}}{\ssr{1}{00'}\s{0'}{2}}}
 \nl
   &&+2F^{(0)}_{[1\,00'\,2]}\left(
  4\Log{\frac{\s{1}{0}}{\ssr{1}{00'}}}\Log{\frac{\s{0'}{2}}{\ssl{00'}{2}}}
  -\frac12\Logsq{\frac{\s{1}{0}\s{0'}{2}\s{1}{2}\s{0}{0'}}{s^2_{1(00')}s^2_{(00')2}}}
-\frac{2\pi^2}{3}+X\right)
\nl 
&&-2\Li_2\left(1-\frac{\s{1}{0}}{\ssr{1}{00'}}\right)
-2\Li_2\left(1-\frac{\s{0'}{2}}{\ssl{00'}{2}}\right)
-2\Li_2\left(1-\frac{\s{1}{2}\s{0}{0'}}{\ssr{1}{00'}\ssl{00'}{2}}\right)+\frac{2\pi^2}{3}
\nl
&& +\Logsq{\frac{\s{1}{0}\ssl{00'}{2}}{\ssr{1}{00'}\s{0'}{2}}}
+\Log{\frac{\s{1}{2}\s{0}{0'}}{\s{1}{0}\s{0'}{2}}}\Log{\frac{\s{1}{0}\s{0'}{2}}{\ssr{1}{00'}\ssl{00'}{2}}},
\label{F2bare}
\ea
with $X=-2\frac{c_\Gamma}{\eps^2}(\Q{1}{00'}{2}^2/\mu^2)^{-\eps}+2\pi^2+O(\eps)$,
and the parity operation $P$: $\{1,0\}\leftrightarrow\{2,0'\}$. 
Here all analytic continuations have been performed, so the
logarithms are all real for timelike (positive) invariants, as is the case for our application.
The infrared divergences are contained in the factor $X$ but in practice all we will need is the fully
renormalized form factor, defined in eq.~(\ref{many_dipoles_subtraction}), which is finite and obtained by a simple substitution:
\be
 F^{(1){\rm ren}}_{[1\,00'\,2]}=F^{(1){\rm bare}}_{[1\,00'\,2]} \quad \mbox{with}\quad
 X\mapsto \frac14\log^2\frac{\z{1}{2}\z{1}{0}\z{0}{0'}}{\zsq{1}{0'}\z{0'}{2}}+\frac14\log^2\frac{\z{1}{2}\z{0}{0'}\z{0'}{2}}{\z{1}{0}\zsq{0}{2}}\,.
 \label{F2ren_appendix}
\ee

\section{Doing transverse integrals efficiently}
\label{app:integrals}

Two-dimensional integrals can be done extremely efficiently with the differential equation method.
Here we elaborate on our implementation, emphasizing the simplifications related to the fact that all the integrals are absolutely convergent
and done directly in $2$ dimensions.  We first illustrate the method on the integral
\be
 g_1(y,\bar{y}) = -\int \frac{d^2z}{\pi} \frac{|1-y|^2}{|1-z|^2|y-z|^2}\log\frac{|y|^2}{|z-1-y|^2|z|^2}\,, \label{defIex}
\ee
which occurs at two-loops when obtaining the translation-invariant kernel $H(y)$ (\ref{translation_invariant_case}).
The idea is to differentiate with respect to $y$ and add a total derivative with respect to $z$ to remove derivatives of rational factors.
Indeed, using the relevant identity:
\be
 \left(\frac{d}{dy}+ \frac{d}{dz} \frac{1-z}{1-y}\right)  \frac{|1-y|^2}{|1-z|^2|y-z|^2} =0,
 \label{inner_derivative_0}
\ee
one readily gets that
\ba
\frac{d}{dy} g_1(y,\bar{y}) &=& -\int \frac{d^2z}{\pi}
\left( \frac{d}{dy} + \frac{d}{dz}\frac{1-z}{1-y}\right)\frac{|1-y|^2}{|1-z|^2|y-z|^2}\log\frac{|y|^2}{|z-1-y|^2|z|^2}\,.
 \label{inner_derivative}
\ea
We ``win'' because the derivatives commutes with the rational factor and hits the logarithm, producing a simpler integral.

An important subtlety is that the left-hand-side of eq.~(\ref{inner_derivative_0})
is singular and so the equation is only strictly valid for generic $z$. There are additional contact terms given
by the ``holomorphic anomaly''
\be
 \frac{d}{dy} \frac{1}{\bar{y}-\bar{z}} =  \frac{d}{d\bar{y}} \frac{1}{y-z} = \pi\delta^2(y{-}z). \label{anomaly}
\ee
This can be understood from the two-dimensional Poisson equation $\partial_z\partial_{\bar z}\log(z\bar{z})=\pi\delta^2(z)$.
These terms would be absent in dimensional regularization but appear because we insist to work with $\eps=0$ (see \cite{Caron-Huot:2014lda} for four-dimensional examples).  In the example (\ref{inner_derivative}), the contact terms are at $z=1$ and $z=y$ but the logarithm turns out to vanish on both, so these can be dropped. Evaluating the derivative then gives simply
\be
 \frac{d}{dy} g_1(y,\bar{y}) = \frac{1}{y} \int \frac{d^2z}{\pi}\frac{(1+y)}{z(1+y-z)}
 \frac{(1-\bar{y})}{(1-\bar{z})(\bar{y}-\bar{z})}\,. \label{derivative_I1}
\ee
To finish, one can repeat the same procedure,
inserting a variant of eq.~(\ref{inner_derivative_0}) to differentiate the integral with respect to $y$ (and/or $\bar{y}$).
Now only the contact term contributes and a general result obtained this way is
\be
 I_{ab,cd} = \int \frac{d^2z}{\pi} \frac{(a-b)}{(z-a)(z-b)}\frac{(\bar{c}-\bar{d})}{(\bar{z}-\bar{c})(\bar{z}-\bar{d})}= \log \frac{|a{-}d|^2|b{-}c|^2}{|a{-}c|^2|b{-}d|^2}\,, \label{Iabcd}
\ee
which gives (using the vanishing at $y=-1$ to fix the integration constant)
\be
 \frac{d}{dy} g_1(y,\bar{y}) = \frac{1}{y} \times 2\log(y\bar{y})\quad \longrightarrow \quad g_1(y,\bar y) = \log^2(y\bar{y})\,. \label{result_g1}
\ee
This result can be easily confirmed by numerical integration.

A critical point to emphasize is that the factor $1/y$ in eq.~(\ref{derivative_I1}) had to be pulled out in front of the integral
before taking the second derivative.  If the derivative were allowed to act on that factor, one would gain nothing from it.
Only properly normalized integrals simplify upon taking derivatives.

A simple criterion to identify properly normalized integrals is that all the Poincar\'e residues
of their rational factors should be constant (these are often called \emph{leading singularities}).
These are simply the double residue with respect to $z$ followed by $\bar{z}$,
of the rational factors in the integrand, with $z$ and $\bar{z}$ treated as independent complex variables.
This property is easily verified in eqs.~(\ref{defIex}) and (\ref{Iabcd}).
Its significance is that it ensures that derivatives of the rational factors have vanishing Poincar\'e residues, which is needed for them
to be total derivatives which simplify upon integration by parts as in eq.~(\ref{inner_derivative_0}).
See for instance refs.~\cite{Henn:2013pwa,Caron-Huot:2014lda} for other applications of this criterion.

The procedure to decompose an integral into properly normalized ones is essentially partial fractions.
When the denominators do not couple $z$ and $\bar{z}$, it is in fact literally partial fraction in these two variables, one after the other.
But the integrals we need also contain in the denominator an irreducible quadratic form $Q(z,\bar{z})$, which is harder to partial-fraction out.
We illustrate this with the other integral appearing in $H^{(2)}(y)$, coming from the second term in the kernel (\ref{K2real}):
\be
\int \frac{d^2z}{\pi}\frac{1}{|1{-}z|^2|y{-}z|^2}\left[ 1+\frac{|y|^2}{Q}\right]\log\frac{|1{-}z|^2|y{-}z|^2}{|1{+}y{-}z|^2|z|^2},\qquad
 Q=|1{-}z|^2|y{-}z|^2-|1{+}y{-}z|^2|z|^2\,.\nonumber
\ee
Despite appearances, $Q$ is a quadratic form in $z, \bar{z}$.
The leading singularities of the rational factor can be computed and found to be linear combinations of $1/[(1{-}y)(1{+}\bar{y})]$ and $1/[(1{+}y)(1{-}\bar{y})]$,
so the decomposition into properly normalized integrals will require two terms.
To illustrate the result of the partial-fraction method to be detailed shortly, one indeed finds that
the integral can be rewritten exactly as:
\be
\frac{1}{(1-y)(1+\bar{y})}
\int \frac{d^2z}{\pi}\frac{y(1{-}y)(1{+}\bar{y})}{(1{-}z)(y{-}z)Q}
\log\frac{|1{-}z|^2|y{-}z|^2}{|1{+}y{-}z|^2|z|^2} + (y\leftrightarrow\bar{y})\,. \label{partial_fraction_quad}
\ee
This rewriting of the integrand is a purely algebraic identity. The upshot is that all residues have been pulled out
and the leading singularities of the rational factor inside the integral are only $\pm1$.
One then expects, and finds, that the derivative $d/dy$ of the integral is a total derivatives in $z$ and $\bar{z}$.
One does not need to make any clever \emph{guess} to find this total derivatives: in practice we
simply write down an ansatz with a polynomial numerator in $z$ and $\bar{z}$ and solve for the coefficients.
We obtain for example the identity:
\be
 \left[\frac{d}{dy} +\frac{d}{dz}\frac{1{-}z}{1{-}y}+
 \left(\frac{d}{dz}(2z{-}1{-}y)+\frac{d}{d\bar{z}}(2\bar{z}{-}1{-}\bar{y}
 )\right)\frac{\bar{y}(1{+}y)(1{-}z)(y{-}z)}{y(1{-}y)(y{+}\bar{y})(1{+}y\bar{y})}\right]\frac{y(1{-}y)(1{+}\bar{y})}{(1{-}z)(y{-}z)Q}=0\,,\nonumber
\ee
again up to contact terms arising from the holomorphic anomaly (at $z=y$ and $z=1$). Plugging this into the integral (\ref{partial_fraction_quad})
thus gives its $y$-derivative in terms of contact terms and the simpler integral (\ref{Iabcd}).
The derivative can then be easily integrated, and the integration constant again is fixed by the vanishing at $y=-1$, yielding
the two-loop linearized kernel:
\be
 H^{(2)}(y) =
 \frac{4\log(y\bar{y})^2}{(1-y)(1-\bar{y})} + 8{\rm Re} \frac{\Li_2(y)-\Li_2(\bar{y})+\Li_2(-y\bar{y})-\log(y\bar{y})\log\frac{1-\bar{y}}{1+y\bar{y}}+\frac12\zeta_2}{(1+y)(1-\bar{y})}\,.
\ee
This agrees precisely with the result in eq.~(105) of Balitsky\& Chirilli 0710.4330.

Chief advantages of this method are its speed and uniform applicability.
Indeed, the basic steps (integration by parts and partial fractions) are algebraic and independent of the transcendental weight of the functions being integrated.
That is, the same code we used to do the two-loop integral $H^{(2)}(y)$ as just described, automatically also worked for $H^{(3)}(y)$ and would presumably work at higher orders as well (producing the result as an iterated integral).

To conclude, we elaborate on partial fractions in the presence of the quadratic form $Q$ in the denominator.
The main step is to exploit the geometry to create a complete basis.
For the integrals involving $Q$, there are other singularities on the 8 lines
$z=0,1,y,1{+}y$ and $\bar{z}=0,1,\bar{y},1{+}\bar{y}$, each line intersecting the quadric $Q$ at two points. However the geometry is a bit degenerate
and there are only 8 intersections: $(z,\bar{z})=\{(1,0),\, (0,1),\, (0,\bar{y}),(y,0),\,(1{+}y,1),$
$\, (1,1{+}\bar{y}),\, (y,1{+}\bar{y}),\,(1{+}y,\bar{y})\}$.
A complete basis of rational functions is then obtained by writing 8 objects $\frac{n_a}{(z-a)Q}$ where $a\in\{0,1,y,1{+}y\}$ and the numerators $n_a$
are linear in $\bar{z}$ and chosen to leave only one Poincar\'e residue nonzero (and equal to 1).  A ninth integral $\sqrt{y\bar{y}}/Q$ (accounting for a residue at infinity), together with simpler integrals with only linear denominators, complete the basis.
Once a basis is fixed, partial fraction identities like (\ref{partial_fraction_quad}) follow simply from computing Poincar\'e residues, a fast operation.

The other integrals needed in this paper, involving the triple-real functions (\ref{subtraction_change}) and (\ref{3looplinear}), were dealt with in a similar way,
although their geometry is somewhat simpler (no square roots appeared in these cases).

\subsection{Single-valued functions for the linearized kernel}

The linearized kernel (\ref{3looplinear}) is a weight 3 function of one complex variable;
the preceding method produces it in the form of an iterated integral, whose integration constants could be easily fixed from the limit $x\to 1$.
Its symbol turns out to be made of the five letters $x,1{-}x,\xb,1{-}\xb$ and $1{-}v=x{+}\xb{-}x\xb$.
At transcendental weight 3, we found only three nontrivial single-valued functions with such symbol, in terms of which
the three-loop linearized kernel $K^{(3){\rm lin}}(x)$ in (\ref{3looplinear_final}) is compactly written:
\begin{subequations}\ba
 O_1&=& 2\big(\Li_3(x)+\Li_3(\xb)-2\zeta_3\big)-\log u\big(\Li_2(x)+\Li_2(\xb)\big), \\
 O_2&=& 2\big(\Li_3(1{-}x)+\Li_3(1{-}\xb)-2\zeta_3\big)-\log v\big(\Li_2(1{-}x)+\Li_2(1{-}\xb)\big),\\
 O_3&=& \left\{\Li_3\left(\frac{\xb}{x(\xb{-}1)}\right)+\Li_3\left(\frac{x(\xb{-}1)}{\xb}\right)
 +\frac12\left[\Li_2\left(\frac{\xb}{x(\xb{-}1)}\right)-\Li_2\left(\frac{x(\xb{-}1)}{\xb}\right)\right]\log (1{-}x)(1{-}\bar{x})\right.
\nl && -4\Li_3(x)-2\Li_3(1{-}x)+\log(x\bar{x})\Li_2(x)+\frac16\log^3(1{-}x)-\frac12\log^2(1{-}x)\big(\!\log(x)-\log(\bar{x})\big)
\nl&& \left.-\frac14\log^2(1{-}x)\log (1{-}x)(1{-}\bar{x})+\zeta_2\log(1{-}x)\right\} -(x\leftrightarrow\xb).
\ea\label{app:O123}\end{subequations}
Although not manifest from these formulas, these functions have no branch cut on the complex plane where $\xb=x^*$.
This can be confirmed by series-expanding around singular points such as $x=\xb=0$ and $x=\xb=1$,
where to all orders one finds only single-valued logarithms of $\log(x\xb)$ or $\log(1{-}x)(1{-}\xb)$ but never $\log(x)$ nor $\log(1{-}x)$ separately.
Furthermore, there are no singularities along $1{-}v=0$ (which traces a unit circle with center at $x=1$).

We note that the same five letters are also singularities of the two-loop kernel, so
it is natural to conjecture that no other letters appear in $K^{(L){\rm lin}}(x)$ to any order in perturbation theory in planar $\mathcal{N}=4$ SYM.
Its translation-invariant projection $H^{(L)}(y)$, defined by the integration (\ref{translation_invariant_case}), can then be obtained by applying the
algorithm detailed above, which implies that at most the ten letters $d\log\{y,\bar{y},1{\pm}y,1{\pm}\bar{y},y{+}\bar{y},1{+}y\bar{y}$,
$\frac{\sqrt{y}{+}i\sqrt{\bar{y}}}{\sqrt{y}{-}i\sqrt{\bar{y}}},\frac{1{+}i\sqrt{y\bar{y}}}{1{-}i\sqrt{y\bar{y}}}\}$ can appear in its symbol (all of which do indeed appear at three loops).

\section{Eigenvalue in terms of harmonic sums for $m=0$ and $m=1$}
\label{app:eigen}

Here we give explicit expressions for the 3-loop Pomeron trajectory, given in coordinate space in eq.~(\ref{3looplinear_final}),
in Mellin space using the harmonic sums
\be
 S_{a}(N) = \sum_{i=1}^N \frac{({\rm sign}\,a)^i}{i^{|a|}},\qquad S_{a_1,\ldots,a_n}(N) = \sum_{i=1}^N \frac{({\rm sign}\,a)^i}{i^{|a|}} S_{a_2,\ldots,a_n}(i)\,.
\ee
This defines the sums for integer $N$ and the Mellin transform produces their analytical continuation from even $N$.
Using standard algorithms \cite{Remiddi:1999ew}, we have converted the Mellin integral projected onto transverse angular momentum
$m=0$, eq.~(\ref{F0Mellin}), to harmonic sums with argument $N=\frac{-1+i\nu}{2}$:
\def\S#1{S_{#1}}
\ba
F^{(1)}_{0,\nu} &=& -4S_1,\qquad F^{(2)}_{0,\nu} =8\S{3}-16\S{-2,1}+8\zeta_2\big(3\S{-1}+3\log2+\S{1}\big)-6\zeta_3,
\\
\frac{F^{(3)}_{0,\nu}}{32} &=& -\S{5} + 2\S{-4, 1} -\S{-3, 2} + 2\S{-2, 3} -\S{2, -3} - 2\S{3, -2}+4\S{-3, 1, 1} +  4\S{1, -3, 1} + 2\S{1, -2, 2} \nl &&
+ 2\S{1, 2, -2} + 2\S{2, 1, -2} - 8\S{1, -2, 1, 1}+\zeta_{2} \big(\S{1}\S{2}-3\S{-3} +2\S{-2,1}-4\S{1,-2}\big)-\tfrac{49}{2}\zeta_4\S{1}
\nl &&
+7\zeta_3\big(2\S{1,-1}+2(\S{1}-\S{-1})\log2-\S{-2}-\log^22\big)+(8\zeta_{-3,1}-17\zeta_4)\big(\S{-1}-\S{1}+\log2\big)
\nl && -\tfrac12\zeta_3\S{2}+4\zeta_5-6\zeta_2\zeta_3+8\zeta_{-3,1,1}\,. \label{F0mellin1}
\ea
Here $\zeta_{-3,1}\approx0.087786 $ and $\zeta_{-3,1,1}\approx-0.009602$ are multi-zeta values.
This result is in precise agreement with \cite{Gromov:2015vua}. The Pomeron trajectory is the sum of $F_{m,\nu}$ and $F_{m,-\nu}$,
see eq.~(\ref{j_from_F}).
For $m\neq 0$ our result is new.
For $m=1$, for example, the Mellin transform can be expressed in terms of harmonic sums now
with argument $N=\frac{i\nu}{2}$, giving the Odderon Regge trajectory:
\ba
F^{(1)}_{1,\nu} &=& -4S_1,\qquad \frac{F^{(2)}_{1,\nu}}{8}=N^{-1}(\S{-2}+\zeta_2)-N^{-2}\S{1}+\S{3}+\zeta_2\S{1}+\tfrac12\zeta_3,\\
\frac{F^{(3)}_{1,\nu}}{16}&=&
N^{-1}\left(-3\S{-4}+2\S{-3,1}+2\S{-2,2}+2\S{1,-3}+4\S{2,-2}-8\S{-2,1,1}+4\S{1,-2,1}-8\S{1,1,-2}\right)
\nl &&
+N^{-2}\big(2\S{3}-\S{-3}-2\S{-2,1}+4\S{1,-2}+4\zeta_2\S{1}-5\zeta_3\big)
+N^{-3}\left(4\S{1,1}-4\S{-2}-\S{2}-3\zeta_2\right)
\nl &&
+N^{-1}\left(\zeta_2(-2\S{1}^2-6\S{-2})+\zeta_3(7\S{-1}+3\S{1})-9\zeta_4\right)
+(3N^{-4}-\tfrac{11}{2}\zeta_4)\S{1}-2\S{5}\nl && -\zeta_2\zeta_3-3\zeta_5\,.
\ea
This is regular and in fact vanishes at $\nu=0$, in accordance with the all-order result (\ref{odderon_intercept}).
Other values of $m$ can be evaluated numerically using the attached Mathematica notebook.

\end{appendix}

\bibliographystyle{JHEP}
\bibliography{regge}

\end{document}

%% file: densitymatrix.pdf_tex
\begingroup%
  \makeatletter%
  \providecommand\color[2][]{%
    \errmessage{(Inkscape) Color is used for the text in Inkscape, but the package 'color.sty' is not loaded}%
    \renewcommand\color[2][]{}%
  }%
  \providecommand\transparent[1]{%
    \errmessage{(Inkscape) Transparency is used (non-zero) for the text in Inkscape, but the package 'transparent.sty' is not loaded}%
    \renewcommand\transparent[1]{}%
  }%
  \providecommand\rotatebox[2]{#2}%
  \ifx\svgwidth\undefined%
    \setlength{\unitlength}{205.59615665bp}%
    \ifx\svgscale\undefined%
      \relax%
    \else%
      \setlength{\unitlength}{\unitlength * \real{\svgscale}}%
    \fi%
  \else%
    \setlength{\unitlength}{\svgwidth}%
  \fi%
  \global\let\svgwidth\undefined%
  \global\let\svgscale\undefined%
  \makeatother%
  \begin{picture}(1,0.32809979)%
    \put(0.44385307,0.34483054){\color[rgb]{0,0,0}\makebox(0,0)[lb]{\smash{}}}%
    \put(0.44464719,0.28714055){\color[rgb]{0,0,0}\makebox(0,0)[lb]{\smash{$U(\theta_1)$}}}%
    \put(0.44413842,0.16124547){\color[rgb]{0,0,0}\makebox(0,0)[lb]{\smash{$U(\theta_0)$}}}%
    \put(0.44534199,0.02696455){\color[rgb]{0,0,0}\makebox(0,0)[lb]{\smash{$U(\theta_2)$}}}%
    \put(0.24999176,-0.01509837){\color[rgb]{0,0,0}\makebox(0,0)[lb]{\smash{}}}%
    \put(0.44248391,0.31338557){\color[rgb]{0,0,0}\makebox(0,0)[lb]{\smash{*}}}%
    \put(0,0){\includegraphics[width=\unitlength,page=1]{densitymatrix.pdf}}%
  \end{picture}%
\endgroup%

%% file: dipole.bbl
\providecommand{\href}[2]{#2}\begingroup\raggedright\begin{thebibliography}{10}

\bibitem{Kuraev:1977fs}
E.~Kuraev, L.~Lipatov, and V.~S. Fadin, {\it {The Pomeranchuk Singularity in
  Nonabelian Gauge Theories}},  {\em Sov.Phys.JETP} {\bf 45} (1977) 199--204.

\bibitem{Balitsky:1978ic}
I.~Balitsky and L.~Lipatov, {\it {The Pomeranchuk Singularity in Quantum
  Chromodynamics}},  {\em Sov.J.Nucl.Phys.} {\bf 28} (1978) 822--829.

\bibitem{Balitsky:1995ub}
I.~Balitsky, {\it {Operator expansion for high-energy scattering}},  {\em
  Nucl.Phys.} {\bf B463} (1996) 99--160,
  [\href{http://xxx.lanl.gov/abs/hep-ph/9509348}{{\tt hep-ph/9509348}}].

\bibitem{Kovchegov:1999yj}
Y.~V. Kovchegov, {\it {Small x F(2) structure function of a nucleus including
  multiple pomeron exchanges}},  {\em Phys.Rev.} {\bf D60} (1999) 034008,
  [\href{http://xxx.lanl.gov/abs/hep-ph/9901281}{{\tt hep-ph/9901281}}].

\bibitem{McLerran:1993ni}
L.~D. McLerran and R.~Venugopalan, {\it {Computing quark and gluon distribution
  functions for very large nuclei}},  {\em Phys. Rev.} {\bf D49} (1994)
  2233--2241, [\href{http://xxx.lanl.gov/abs/hep-ph/9309289}{{\tt
  hep-ph/9309289}}].

\bibitem{Gelis:2010nm}
F.~Gelis, E.~Iancu, J.~Jalilian-Marian, and R.~Venugopalan, {\it {The Color
  Glass Condensate}},  {\em Ann. Rev. Nucl. Part. Sci.} {\bf 60} (2010)
  463--489, [\href{http://xxx.lanl.gov/abs/1002.0333}{{\tt arXiv:1002.0333}}].

\bibitem{Caporale:2015vya}
F.~Caporale, G.~Chachamis, B.~Murdaca, and A.~S. Vera, {\it
  {Balitsky-Fadin-Kuraev-Lipatov Predictions for Inclusive Three Jet Production
  at the LHC}},  {\em Phys. Rev. Lett.} {\bf 116} (2016), no.~1 012001,
  [\href{http://xxx.lanl.gov/abs/1508.0771}{{\tt arXiv:1508.0771}}].

\bibitem{Iancu:2015joa}
E.~Iancu, J.~D. Madrigal, A.~H. Mueller, G.~Soyez, and D.~N.
  Triantafyllopoulos, {\it {Collinearly-improved BK evolution meets the HERA
  data}},  {\em Phys. Lett.} {\bf B750} (2015) 643--652,
  [\href{http://xxx.lanl.gov/abs/1507.0365}{{\tt arXiv:1507.0365}}].

\bibitem{Balitsky:2008zza}
I.~Balitsky and G.~A. Chirilli, {\it {Next-to-leading order evolution of color
  dipoles}},  {\em Phys.Rev.} {\bf D77} (2008) 014019,
  [\href{http://xxx.lanl.gov/abs/0710.4330}{{\tt arXiv:0710.4330}}].

\bibitem{Fadin:1998py}
V.~S. Fadin and L.~Lipatov, {\it {BFKL pomeron in the next-to-leading
  approximation}},  {\em Phys.Lett.} {\bf B429} (1998) 127--134,
  [\href{http://xxx.lanl.gov/abs/hep-ph/9802290}{{\tt hep-ph/9802290}}].

\bibitem{Ciafaloni:1998gs}
M.~Ciafaloni and G.~Camici, {\it {Energy scale(s) and next-to-leading BFKL
  equation}},  {\em Phys. Lett.} {\bf B430} (1998) 349--354,
  [\href{http://xxx.lanl.gov/abs/hep-ph/9803389}{{\tt hep-ph/9803389}}].

\bibitem{Salam:1998tj}
G.~P. Salam, {\it {A Resummation of large subleading corrections at small x}},
  {\em JHEP} {\bf 07} (1998) 019,
  [\href{http://xxx.lanl.gov/abs/hep-ph/9806482}{{\tt hep-ph/9806482}}].

\bibitem{Vera:2005jt}
A.~Sabio~Vera, {\it {An 'All-poles' approximation to collinear resummations in
  the Regge limit of perturbative QCD}},  {\em Nucl. Phys.} {\bf B722} (2005)
  65--80, [\href{http://xxx.lanl.gov/abs/hep-ph/0505128}{{\tt
  hep-ph/0505128}}].

\bibitem{Iancu:2015vea}
E.~Iancu, J.~D. Madrigal, A.~H. Mueller, G.~Soyez, and D.~N.
  Triantafyllopoulos, {\it {Resumming double logarithms in the QCD evolution of
  color dipoles}},  {\em Phys. Lett.} {\bf B744} (2015) 293--302,
  [\href{http://xxx.lanl.gov/abs/1502.0564}{{\tt arXiv:1502.0564}}].

\bibitem{Gromov:2015vua}
N.~Gromov, F.~Levkovich-Maslyuk, and G.~Sizov, {\it {Pomeron Eigenvalue at
  Three Loops in $\mathcal N=$ 4 Supersymmetric Yang-Mills Theory}},  {\em
  Phys. Rev. Lett.} {\bf 115} (2015), no.~25 251601,
  [\href{http://xxx.lanl.gov/abs/1507.0401}{{\tt arXiv:1507.0401}}].

\bibitem{Velizhanin:2015xsa}
V.~N. Velizhanin, {\it {BFKL pomeron in the next-to-next-to-leading
  approximation in the planar N=4 SYM theory}},
  \href{http://xxx.lanl.gov/abs/1508.0285}{{\tt arXiv:1508.0285}}.

\bibitem{Balitsky:2015tca}
I.~Balitsky, V.~Kazakov, and E.~Sobko, {\it {Three-point correlator of twist-2
  operators in BFKL limit}},  \href{http://xxx.lanl.gov/abs/1506.0203}{{\tt
  arXiv:1506.0203}}.

\bibitem{Brower:2006ea}
R.~C. Brower, J.~Polchinski, M.~J. Strassler, and C.-I. Tan, {\it {The Pomeron
  and gauge/string duality}},  {\em JHEP} {\bf 0712} (2007) 005,
  [\href{http://xxx.lanl.gov/abs/hep-th/0603115}{{\tt hep-th/0603115}}].

\bibitem{Caron-Huot:2013fea}
S.~Caron-Huot, {\it {When does the gluon reggeize?}},  {\em JHEP} {\bf 05}
  (2015) 093, [\href{http://xxx.lanl.gov/abs/1309.6521}{{\tt
  arXiv:1309.6521}}].

\bibitem{Banfi:2002hw}
A.~Banfi, G.~Marchesini, and G.~Smye, {\it {Away from jet energy flow}},  {\em
  JHEP} {\bf 08} (2002) 006,
  [\href{http://xxx.lanl.gov/abs/hep-ph/0206076}{{\tt hep-ph/0206076}}].

\bibitem{Marchesini:2015ica}
G.~Marchesini and A.~H. Mueller, {\it {The BMS Equation and $c\bar{c}$
  Production; A Comparison of the BMS and BK Equations}},  {\em JHEP} {\bf 02}
  (2016) 010, [\href{http://xxx.lanl.gov/abs/1510.0876}{{\tt
  arXiv:1510.0876}}].

\bibitem{Hatta:2008st}
Y.~Hatta, {\it {Relating e+ e- annihilation to high energy scattering at weak
  and strong coupling}},  {\em JHEP} {\bf 0811} (2008) 057,
  [\href{http://xxx.lanl.gov/abs/0810.0889}{{\tt arXiv:0810.0889}}].

\bibitem{Caron-Huot:2015bja}
S.~Caron-Huot, {\it {Resummation of non-global logarithms and the BFKL
  equation}},  \href{http://xxx.lanl.gov/abs/1501.0375}{{\tt arXiv:1501.0375}}.

\bibitem{Hofman:2008ar}
D.~M. Hofman and J.~Maldacena, {\it {Conformal collider physics: Energy and
  charge correlations}},  {\em JHEP} {\bf 0805} (2008) 012,
  [\href{http://xxx.lanl.gov/abs/0803.1467}{{\tt arXiv:0803.1467}}].

\bibitem{Cornalba:2009ax}
L.~Cornalba, M.~S. Costa, and J.~Penedones, {\it {Deep Inelastic Scattering in
  Conformal QCD}},  {\em JHEP} {\bf 03} (2010) 133,
  [\href{http://xxx.lanl.gov/abs/0911.0043}{{\tt arXiv:0911.0043}}].

\bibitem{Weigert:2003mm}
H.~Weigert, {\it {Nonglobal jet evolution at finite N(c)}},  {\em Nucl. Phys.}
  {\bf B685} (2004) 321--350,
  [\href{http://xxx.lanl.gov/abs/hep-ph/0312050}{{\tt hep-ph/0312050}}].

\bibitem{Nagy:2012bt}
Z.~Nagy and D.~E. Soper, {\it {Parton shower evolution with subleading color}},
   {\em JHEP} {\bf 1206} (2012) 044,
  [\href{http://xxx.lanl.gov/abs/1202.4496}{{\tt arXiv:1202.4496}}].

\bibitem{Larkoski:2015zka}
A.~J. Larkoski, I.~Moult, and D.~Neill, {\it {Non-Global Logarithms,
  Factorization, and the Soft Substructure of Jets}},  {\em JHEP} {\bf 09}
  (2015) 143, [\href{http://xxx.lanl.gov/abs/1501.0459}{{\tt
  arXiv:1501.0459}}].

\bibitem{Becher:2015hka}
T.~Becher, M.~Neubert, L.~Rothen, and D.~Y. Shao, {\it {Effective Field Theory
  for Jet Processes}},  {\em Phys. Rev. Lett.} {\bf 116} (2016), no.~19 192001,
  [\href{http://xxx.lanl.gov/abs/1508.0664}{{\tt arXiv:1508.0664}}].

\bibitem{Catani:1999ss}
S.~Catani and M.~Grazzini, {\it {Infrared factorization of tree level QCD
  amplitudes at the next-to-next-to-leading order and beyond}},  {\em
  Nucl.Phys.} {\bf B570} (2000) 287--325,
  [\href{http://xxx.lanl.gov/abs/hep-ph/9908523}{{\tt hep-ph/9908523}}].

\bibitem{Feige:2014wja}
I.~Feige and M.~D. Schwartz, {\it {Hard-Soft-Collinear Factorization to All
  Orders}},  {\em Phys.Rev.} {\bf D90} (2014), no.~10 105020,
  [\href{http://xxx.lanl.gov/abs/1403.6472}{{\tt arXiv:1403.6472}}].

\bibitem{Balitsky:2009xg}
I.~Balitsky and G.~A. Chirilli, {\it {NLO evolution of color dipoles in N=4
  SYM}},  {\em Nucl.Phys.} {\bf B822} (2009) 45--87,
  [\href{http://xxx.lanl.gov/abs/0903.5326}{{\tt arXiv:0903.5326}}].

\bibitem{Avsar:2009yb}
E.~Avsar, Y.~Hatta, and T.~Matsuo, {\it {Soft gluons away from jets:
  Distribution and correlation}},  {\em JHEP} {\bf 0906} (2009) 011,
  [\href{http://xxx.lanl.gov/abs/0903.4285}{{\tt arXiv:0903.4285}}].

\bibitem{Angeles-Martinez:2016dph}
R.~Ángeles Martínez, J.~R. Forshaw, and M.~H. Seymour, {\it {Ordering multiple
  soft gluon emissions}},  {\em Phys. Rev. Lett.} {\bf 116} (2016), no.~21
  212003, [\href{http://xxx.lanl.gov/abs/1602.0062}{{\tt arXiv:1602.0062}}].

\bibitem{Kinoshita:1962ur}
T.~Kinoshita, {\it {Mass singularities of Feynman amplitudes}},  {\em
  J.Math.Phys.} {\bf 3} (1962) 650--677.

\bibitem{Lee:1964is}
T.~Lee and M.~Nauenberg, {\it {Degenerate Systems and Mass Singularities}},
  {\em Phys.Rev.} {\bf 133} (1964) B1549--B1562.

\bibitem{Bern:1999ry}
Z.~Bern, V.~Del~Duca, W.~B. Kilgore, and C.~R. Schmidt, {\it {The infrared
  behavior of one loop QCD amplitudes at next-to-next-to leading order}},  {\em
  Phys.Rev.} {\bf D60} (1999) 116001,
  [\href{http://xxx.lanl.gov/abs/hep-ph/9903516}{{\tt hep-ph/9903516}}].

\bibitem{Catani:2000pi}
S.~Catani and M.~Grazzini, {\it {The soft gluon current at one loop order}},
  {\em Nucl.Phys.} {\bf B591} (2000) 435--454,
  [\href{http://xxx.lanl.gov/abs/hep-ph/0007142}{{\tt hep-ph/0007142}}].

\bibitem{Bern:1997nh}
Z.~Bern, J.~S. Rozowsky, and B.~Yan, {\it {Two loop four gluon amplitudes in
  N=4 superYang-Mills}},  {\em Phys. Lett.} {\bf B401} (1997) 273--282,
  [\href{http://xxx.lanl.gov/abs/hep-ph/9702424}{{\tt hep-ph/9702424}}].

\bibitem{Bourjaily:2011hi}
J.~L. Bourjaily, A.~DiRe, A.~Shaikh, M.~Spradlin, and A.~Volovich, {\it {The
  Soft-Collinear Bootstrap: N=4 Yang-Mills Amplitudes at Six and Seven Loops}},
   {\em JHEP} {\bf 03} (2012) 032,
  [\href{http://xxx.lanl.gov/abs/1112.6432}{{\tt arXiv:1112.6432}}].

\bibitem{Bourjaily:2015bpz}
J.~L. Bourjaily, P.~Heslop, and V.-V. Tran, {\it {Perturbation Theory at Eight
  Loops: Novel Structures and the Breakdown of Manifest Conformality}},
  \href{http://xxx.lanl.gov/abs/1512.0791}{{\tt arXiv:1512.0791}}.

\bibitem{Bourjaily:2010wh}
J.~L. Bourjaily, {\it {Efficient Tree-Amplitudes in N=4: Automatic BCFW
  Recursion in Mathematica}},  \href{http://xxx.lanl.gov/abs/1011.2447}{{\tt
  arXiv:1011.2447}}.

\bibitem{Bern:1994zx}
Z.~Bern, L.~J. Dixon, D.~C. Dunbar, and D.~A. Kosower, {\it {One loop n point
  gauge theory amplitudes, unitarity and collinear limits}},  {\em Nucl. Phys.}
  {\bf B425} (1994) 217--260,
  [\href{http://xxx.lanl.gov/abs/hep-ph/9403226}{{\tt hep-ph/9403226}}].

\bibitem{Bern:1994cg}
Z.~Bern, L.~J. Dixon, D.~C. Dunbar, and D.~A. Kosower, {\it {Fusing gauge
  theory tree amplitudes into loop amplitudes}},  {\em Nucl. Phys.} {\bf B435}
  (1995) 59--101, [\href{http://xxx.lanl.gov/abs/hep-ph/9409265}{{\tt
  hep-ph/9409265}}].

\bibitem{Drummond:2008vq}
J.~M. Drummond, J.~Henn, G.~P. Korchemsky, and E.~Sokatchev, {\it {Dual
  superconformal symmetry of scattering amplitudes in N=4 super-Yang-Mills
  theory}},  {\em Nucl. Phys.} {\bf B828} (2010) 317--374,
  [\href{http://xxx.lanl.gov/abs/0807.1095}{{\tt arXiv:0807.1095}}].

\bibitem{Bourjaily:2013mma}
J.~L. Bourjaily, S.~Caron-Huot, and J.~Trnka, {\it {Dual-Conformal
  Regularization of Infrared Loop Divergences and the Chiral Box Expansion}},
  {\em JHEP} {\bf 01} (2015) 001,
  [\href{http://xxx.lanl.gov/abs/1303.4734}{{\tt arXiv:1303.4734}}].

\bibitem{Bern:2005iz}
Z.~Bern, L.~J. Dixon, and V.~A. Smirnov, {\it {Iteration of planar amplitudes
  in maximally supersymmetric Yang-Mills theory at three loops and beyond}},
  {\em Phys.Rev.} {\bf D72} (2005) 085001,
  [\href{http://xxx.lanl.gov/abs/hep-th/0505205}{{\tt hep-th/0505205}}].

\bibitem{Beisert:2006ez}
N.~Beisert, B.~Eden, and M.~Staudacher, {\it {Transcendentality and Crossing}},
   {\em J.Stat.Mech.} {\bf 0701} (2007) P01021,
  [\href{http://xxx.lanl.gov/abs/hep-th/0610251}{{\tt hep-th/0610251}}].

\bibitem{Lipatov:1985uk}
L.~Lipatov, {\it {The Bare Pomeron in Quantum Chromodynamics}},  {\em
  Sov.Phys.JETP} {\bf 63} (1986) 904--912.

\bibitem{Kovchegov:2012rz}
Y.~V. Kovchegov, {\it {Running Coupling Evolution for Diffractive Dissociation
  and the NLO Odderon Intercept}},  {\em AIP Conf.Proc.} {\bf 1523} (2012)
  335--338, [\href{http://xxx.lanl.gov/abs/1212.2113}{{\tt arXiv:1212.2113}}].

\bibitem{Kotikov:1990kg}
A.~V. Kotikov, {\it {Differential equations method: New technique for massive
  Feynman diagrams calculation}},  {\em Phys. Lett.} {\bf B254} (1991)
  158--164.

\bibitem{Gehrmann:1999as}
T.~Gehrmann and E.~Remiddi, {\it {Differential equations for two loop four
  point functions}},  {\em Nucl. Phys.} {\bf B580} (2000) 485--518,
  [\href{http://xxx.lanl.gov/abs/hep-ph/9912329}{{\tt hep-ph/9912329}}].

\bibitem{Henn:2013pwa}
J.~M. Henn, {\it {Multiloop integrals in dimensional regularization made
  simple}},  {\em Phys. Rev. Lett.} {\bf 110} (2013) 251601,
  [\href{http://xxx.lanl.gov/abs/1304.1806}{{\tt arXiv:1304.1806}}].

\bibitem{Caron-Huot:2014lda}
S.~Caron-Huot and J.~M. Henn, {\it {Iterative structure of finite loop
  integrals}},  {\em JHEP} {\bf 06} (2014) 114,
  [\href{http://xxx.lanl.gov/abs/1404.2922}{{\tt arXiv:1404.2922}}].

\bibitem{Remiddi:1999ew}
E.~Remiddi and J.~A.~M. Vermaseren, {\it {Harmonic polylogarithms}},  {\em Int.
  J. Mod. Phys.} {\bf A15} (2000) 725--754,
  [\href{http://xxx.lanl.gov/abs/hep-ph/9905237}{{\tt hep-ph/9905237}}].

\bibitem{Maitre:2005uu}
D.~Maitre, {\it {HPL, a mathematica implementation of the harmonic
  polylogarithms}},  {\em Comput. Phys. Commun.} {\bf 174} (2006) 222--240,
  [\href{http://xxx.lanl.gov/abs/hep-ph/0507152}{{\tt hep-ph/0507152}}].

\bibitem{Kotikov:2002ab}
A.~V. Kotikov and L.~N. Lipatov, {\it {DGLAP and BFKL equations in the N=4
  supersymmetric gauge theory}},  {\em Nucl. Phys.} {\bf B661} (2003) 19--61,
  [\href{http://xxx.lanl.gov/abs/hep-ph/0208220}{{\tt hep-ph/0208220}}].
  [Erratum: Nucl. Phys.B685,405(2004)].

\bibitem{Basso:2014pla}
B.~Basso, S.~Caron-Huot, and A.~Sever, {\it {Adjoint BFKL at finite coupling: a
  short-cut from the collinear limit}},  {\em JHEP} {\bf 01} (2015) 027,
  [\href{http://xxx.lanl.gov/abs/1407.3766}{{\tt arXiv:1407.3766}}].

\bibitem{Marzani:2007gk}
S.~Marzani, R.~D. Ball, P.~Falgari, and S.~Forte, {\it {BFKL at
  next-to-next-to-leading order}},  {\em Nucl. Phys.} {\bf B783} (2007)
  143--175, [\href{http://xxx.lanl.gov/abs/0704.2404}{{\tt arXiv:0704.2404}}].

\bibitem{Costa:2012cb}
M.~S. Costa, V.~Goncalves, and J.~Penedones, {\it {Conformal Regge theory}},
  {\em JHEP} {\bf 12} (2012) 091,
  [\href{http://xxx.lanl.gov/abs/1209.4355}{{\tt arXiv:1209.4355}}].

\bibitem{Basso:2015uxa}
B.~Basso, A.~Sever, and P.~Vieira, {\it {Hexagonal Wilson Loops in Planar
  $\mathcal{N}=4$ SYM Theory at Finite Coupling}},
  \href{http://xxx.lanl.gov/abs/1508.0304}{{\tt arXiv:1508.0304}}.

\bibitem{Balitsky:2013fea}
I.~Balitsky and G.~A. Chirilli, {\it {Rapidity evolution of Wilson lines at the
  next-to-leading order}},  {\em Phys.Rev.} {\bf D88} (2013) 111501,
  [\href{http://xxx.lanl.gov/abs/1309.7644}{{\tt arXiv:1309.7644}}].

\bibitem{Kovner:2013ona}
A.~Kovner, M.~Lublinsky, and Y.~Mulian, {\it {Jalilian-Marian, Iancu, McLerran,
  Weigert, Leonidov, Kovner evolution at next to leading order}},  {\em
  Phys.Rev.} {\bf D89} (2014) 061704,
  [\href{http://xxx.lanl.gov/abs/1310.0378}{{\tt arXiv:1310.0378}}].

\bibitem{Almelid:2015jia}
{\O}.~Almelid, C.~Duhr, and E.~Gardi, {\it {Three-loop corrections to the soft
  anomalous dimension in multi-leg scattering}},
  \href{http://xxx.lanl.gov/abs/1507.0004}{{\tt arXiv:1507.0004}}.

\end{thebibliography}\endgroup
